\documentclass[9pt,journal,compsoc]{IEEEtran}
\usepackage{amsmath,amsfonts}
\usepackage{algorithmic}
\usepackage{algorithm}
\usepackage{array}
\usepackage[caption=false,font=normalsize,labelfont=sf,textfont=sf]{subfig}
\usepackage{textcomp}
\usepackage{stfloats}
\usepackage{url}
\usepackage{verbatim}
\usepackage{graphicx}
\usepackage{cite}
\usepackage{multirow}
\usepackage{makecell}
\usepackage{amssymb} 
\usepackage{amsmath}
\usepackage{subcaption}
\usepackage{pifont}
\usepackage{caption}
\usepackage[font=scriptsize]{caption} 
\begin{document}


\title{A Transverse-Read-assisted Fast Valid-Bits Collection in \\Stochastic Computing MACs for Energy-Efficient in-RTM DNNs}

\author{Jihe Wang, Zhiying Zhang, Xingwu Dong, and Danghui Wang
\thanks{J. Wang, Z. Zhang, and D. Wang are with the School of Computer Science, Northwestern Polytechnical University and Engineering Research Center of Embedded System Integration, Ministry of Education, Xi’an, Shaanxi 710072, China.}
\thanks{
X. Dong is with AVIC ACTRI, China.}
}

\maketitle

\begin{abstract}
It looks very attractive to coordinate racetrack-memory (RM) and stochastic-computing (SC) jointly to build an ultra-low power neuron-architecture.
However, the above combination has always been questioned in a fatal weakness that the heavy valid-bits collection of RM-MTJ, a.k.a. accumulative parallel
counters (APCs), cannot physically match the requirement for energy-efficient in-memory DNNs.
Fortunately, a recently developed Transverse-Read (TR) provides a lightweight collection of valid-bits by detecting domain-wall resistance between a couple of MTJs on a single nanowire.
In this work, we first propose a neuron-architecture that utilizes parallel TRs to build an ultra-fast valid-bits collection for SC, in which, a vector multiplication is successfully degraded as swift TRs. 
To solve huge storage for full stochastic sequences caused by the limited TR banks, a hybrid coding, pseudo-fractal compression, is designed to generate stochastic sequences by segments.
To overcome the misalignment by the parallel early-termination, an asynchronous schedule of TR is further designed to regularize the vectorization, in which, the valid-bits from different lanes are merged in multiple RM-stacks for vector-level valid-bits collection.
However, an inherent defect of TR, i.e., neighbor parts cannot be accessed simultaneously, could limit the throughput of the parallel vector multiplication, therefore, an interleaving data placement is used for full utilization of memory bus among different vectors.
The results show that the SC-MAC assisted with TR achieves $2.88\times-4.40\times $speedup compared to  CORUSCANT, at the same time, energy consumption is reduced by $1.26\times-1.42\times$.

\end{abstract}


\section{Introduction}
\IEEEPARstart{S}{tochastic} computing (SC), a lightweight computing method, reduces the area and energy consumption by gate-level operations \cite{lpSC1,lpSC2,lpSC3}.
However, current SC solutions are still far from practical application because of the expensive bitwise conversion between binary numbers (BN) and stochastic numbers (SN) \cite{s2bsolu1,s2bsolu2,s2bsolu3}.
The recently developed racetrack memory (RTM) \cite{RTM1,RTM2} is able to provide a transverse read (TR) mechanism \cite{TRop}, that empowers a global view of sequences within memory for a lightweight counting of valid-bits to overcome the bitwise conversion.
Therefore, this work first proposes an in-RTM processing architecture assisted with TR to accelerate SC-operations of DNNs \cite{SC+DNN1, SC+DNN2, DNN1}, in which, the heavy stochastic-to-binary (S2B) conversions can be totally removed.

DNNs deployed in traditional in-memory processing architectures face energy efficiency issues at both the storage and computing sides: 1) The refresh operation of DRAM causes high static power in memory \cite{DRAM-FRESH}; 2) The MAC units processing DNNs need large area and energy consumption in limited resources environment \cite{MAC-HIGHEN1, MAC-HIGHEN2}.
Recently, combining SC and RTM to rebuild a RTM processing architecture becomes a new solution to the high energy efficiency \cite{SC+RTM}.
This new architecture used non-volatile memory with near-zero static power consumption on the memory side and SCs with gate-level operations that can drastically reduce the
overhead of on-chip logic. 
However, in this high energy efficient architecture, the accumulative parallel counter (APC) becomes a new bottleneck of energy and area overhead \cite{APC-HIGH} which converses stochastic sequences to binary numbers by counting '1' bit-by-bit, i.e., valid-bits collection, between two DNN layers \cite{DNN1,DNN2}. 
Our study shows that in SC, APCs consume $65.9\%$ more energy than AND operations. 
Moreover, in highly concurrent systems, the large number of APCs required occupies significant additional space, making it challenging to deploy within near-memory architectures.
Therefore, it is urgent to find a new valid-bits collection for energy-efficient SC-DNNs.

\begin{figure}[!t]
    \centering
    \includegraphics[width=0.8\linewidth]{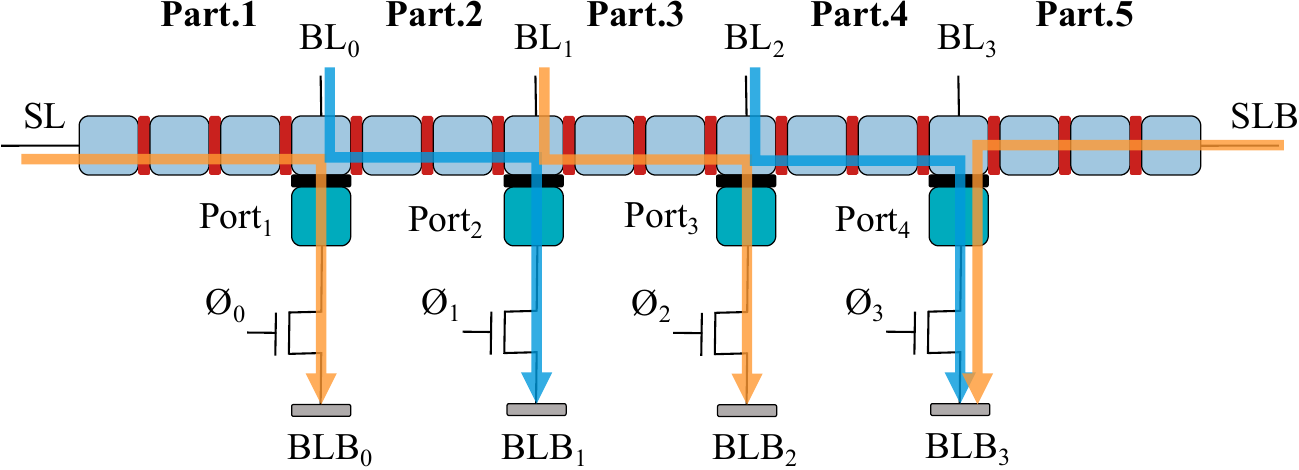}
    \caption{A 16-bit nanowire with four access ports for interval parallel TR \cite{TRop}.}
    \label{Parallel TR}
\end{figure}

Recently, TR technology has been developed in RTMs to provide a lightweight way of valid-bits collection, in which, the valid-bits in a nanowire can be counted by measuring the resistance of domain walls in only one cycle \cite{TRop}.
To increase the parallelism of valid-bits collection, TR only requires adding additional read/write ports along the nanowires which allows TR to read out the bits from each part in parallel as shown in Fig.\ref{Parallel TR}.
TR is promising in solving the bottleneck from APC.
For example, in the case of 8-bit parallel processing, APC-based valid-bits collection requires 8 separate APC units, whereas TR-based valid-bits collection only requires 8 read/write ports on the nanowires. 

However, in order to utilize TR in valid-bits collection for energy-efficient DNNs, there are still three challenges that damage the performance and throughput of parallel SC-MAC: 
\textbf{1)} the limited TR banks force the result stochastic sequences to be temporarily stored in memory buffers, leading to a substantial increase in storage requirements for the memory buffer.
\textbf{2)} the operational misalignment by early-termination leads to a misfortune that, in the sparse vector of DNNs, a majority of fast results have to be latched in registers to wait for a minority of slow ones \cite{DNN1}, bringing about a very low workload of TR;
and \textbf{3)} during TR operation, the voltage detection to a segment incurs an intrinsic defect that the neighbor parts cannot be accessed at the same time, so the TR throughput is only half its theoretical value.
To overcome the above challenges, the MAC assisted with TR should be scrutinized for energy-efficient in-RTM DNNs. 


In this work, we first propose a neuron-architecture that utilizes parallel TRs to build an ultra-fast valid-bits collection for SC, in which, a vector multiplication is successfully degraded as swift TRs. 
Firstly, the Pseudo-Fractal Compression (PFC) is proposed to express the repeatable part of a SN and only generate segments on demand to do TR operation.
Secondly, multiple stacks are built in RTM to merge the result segments from different lanes, which adopts ``asynchronous write-in with synchronous TR'' to achieve regularized vector multiplication.
Finally, to hide the inaccessibility of neighbor parts, the data in multiple vectors are scheduled with the interleaving pattern, with which, the memory bus can be fully utilized. 

The results show that our architecture demonstrates a $1.26\times$ to $1.42\times$ reduction in energy consumption and achieves a $2.88\times$ to $4.40\times$ increase in computational speed compared to other processing-in-memory (PIM) architectures, while also maintaining high stochastic accuracy.
In summary, our main contributions are:

$\bullet$ We firstly propose using TR to accelerate the slow valid-bits collection with APC in SC.

$\bullet$ We propose a parallel architecture to provide high throughput SC-MAC units.

$\bullet$ We propose an in-RTM architecture to schedule the task of vector multiplication to achieve high throughput DNNs.

The remainder of this paper is structured as follows. 
Section 2 introduces the bottleneck of APC in SC-MAC and then proposes the challenges to utilize TR-based valid-bits collection for energy-efficient DNNs.
Section 3 introduces the overall architecture of SC-MAC assisted with TR and the detailed design of neuron-architecture.
Section 4 introduces the reconfiguration of the TR parallelism and BN length in our design.
Section 5 reveals several details of our implementation.
Section 6 evaluates the latency, energy, and accuracy advantages of the proposed design compared with other PIM architectures.
Section 7 concludes the whole paper.




\section{background and motivation}
\subsection{The bottleneck of APC in SC-MAC}


Low-dependency SC (LD-SC) offers tremendous area, power-consumption, and powerful fault tolerance for MAC.
By using LD-SC, the traditional APC-based SC-MAC includes B2S conversion, gate operation, S2B conversion and accumulation. 
The binary number is converted to a SN using a stochastic number generator (SNG) \cite{SNG} and another binary number is converted to a UN using a unary number generator (UNG) \cite{SC+DNN2}. 
The next step involves bit-wise parallel AND operations on UN and SN and output in the form of segments which is then fed into the APC for S2B conversion. Finally, the BNs from all dot product units are sent to the accumulation unit to obtain the final result.

However, it is observed that the imbalance in energy consumption results in a significant difference of $65.9\%$ between the energy consumed by valid-bits collection in the APCs and that of the AND operation.
Additionally, the heavy area overhead and energy consumption caused by adding multiple APCs to improve parallelism make it challenging to deploy the MAC unit near memory.
Therefore, it is urgent to seek a lightweight method to count valid-bits instead of using APC.



\subsection{Lightweight valid-bits collection by Transverse Read}
A TR operation is to detect the resistance of two ports of RTM so that the number of several valid-bits is counted in a single cycle as shown in eq.(1).
On a nanowire, due to the resistance caused by the reflection of electrons by the domain wall between '0' and '1', the resistance detected by TR is proportional to the number of '1' in the nanowire.
As shown in Fig.~\ref{Parallel TR}, the TR of Part.2 starts by applying a read current (voltage) from $BL_0$ to $BLB_1$ with $\phi_1$ set and the $BL_0$-$BLB_1$ voltage (current) can be detected as the number of valid-bit between $port_1$ and $port_2$.
To facilitate parallel valid-bits collection, TR can be employed by adding ports along the nanowire as shown in Fig.~\ref{Parallel TR}. 
After writing the SN across these ports into different parts of the nanowire, the valid-bits can be collected in parallel within a single cycle. 
This approach reduces the significant area overhead associated with multiple APCs and could be a promising substitute for heavy APCs.

\begin{equation}
    TR(segment) = \sum_{i=1}^m APC(bit_i)
\end{equation}

\begin{figure}[!t]
    \centering
    \includegraphics[width=0.85\linewidth]{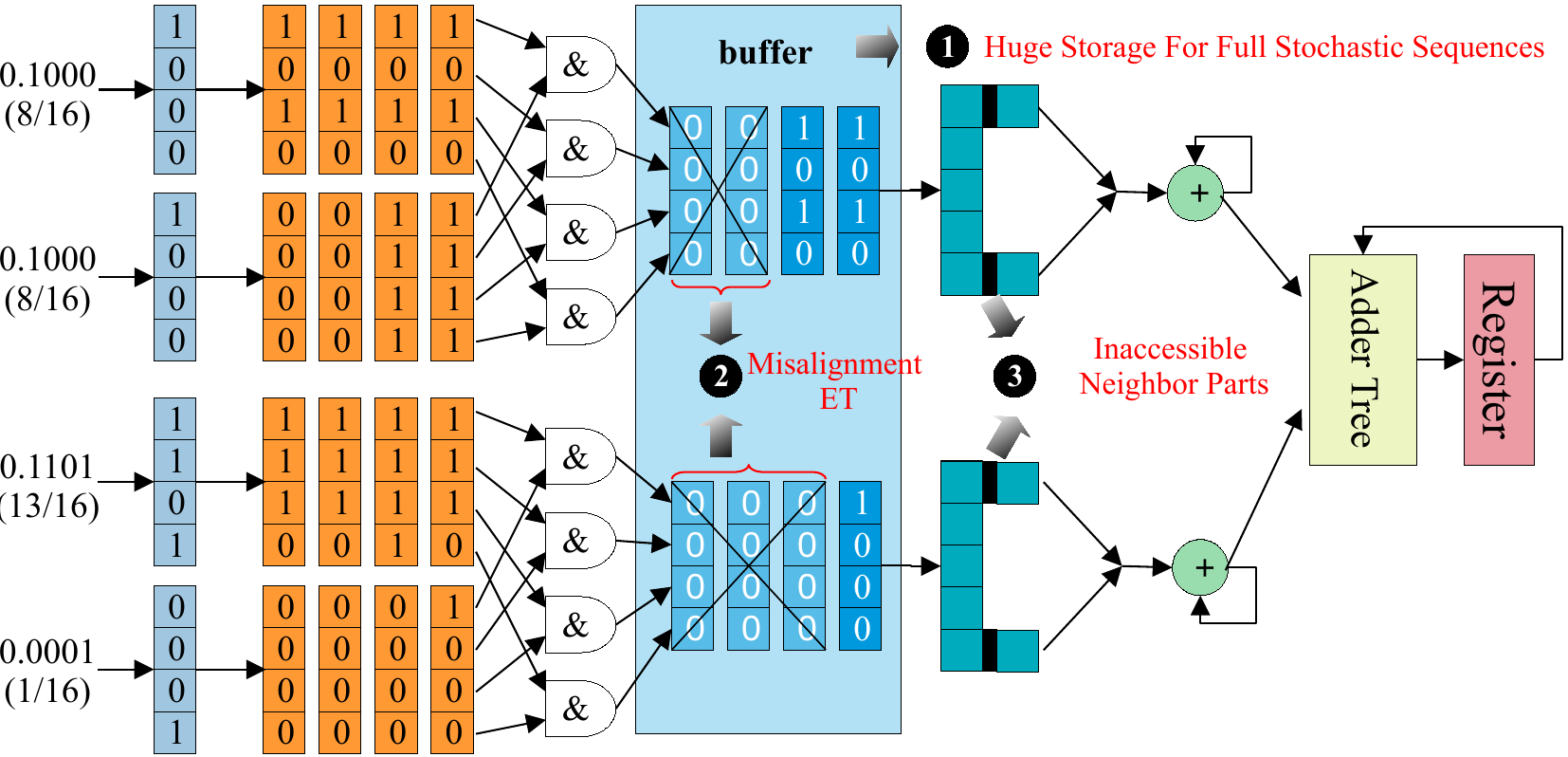}
    \caption{A simple LD-SC MAC that replaces APC with TR. Disadvantage: \ding{182} huge storage for full stochastic sequence; \ding{183} misalignment ET slowing the concurrency of vectors; \ding{184} inaccessible neighbor parts within nanowires for TR.}
    \label{Simple TR}
\end{figure}

To combine TR with LD-SC MAC, the current LD-SC MAC needs to be refactored as shown in Fig.~\ref{Simple TR}.
In the TR + LD-SC MAC architecture, the dot product operation in vector calculation resembles the operation in APC-based LD-SC MAC. However, instead of feeding the results from the dot product unit into APCs one by one, the results stochastic sequences are written into the nanowires through the available write ports.  
Then TRs read bits in nanowires in parallel through the access ports and count the number of '1' for the S2B conversion. 
Finally, the binary results of all dot product units are accumulated.
Compared to APC-based S2B conversion, the energy consumption of TR-based S2B conversion is reduced by $10\times$. 
Additionally, at the memory end, the use of nanowires for near-memory computing through TR operations results in almost no area overhead when compared to APC-based approaches.
This significant reduction in both energy and area overhead makes TR-based S2B conversion a more efficient alternative to the traditional APC method.

However, TR-based refactor still has three challenges.
In this subsection, we analyze the challenges that hinder the high parallelism of the TR-assisted LD-SC MAC, at each end of which, we present individual strategies to address each of these issues. 

\subsection{Challenges and Motivation}
\subsubsection{Heavy storage with full stochastic sequences}  
During the SC computation, the long SN and UN are fed into AND operation and then the result stochastic sequences from all MAC units waiting for the TR operation are stored temporarily in the memory buffer, so long result stochastic sequences greatly increase the storage pressure of the PIM.
Upon comparison, it was found that the space overhead of storing a 256-bit stochastic sequence can store its corresponding 32 8-bit binary numbers, which is the data required for 16 multiplications. 
The reason for this is the low representation efficiency of SC coding, which makes stochastic sequences require an exponential amount of storage space for binary numbers.
Meanwhile, in RTM's shift to access pattern, a 256-bit stochastic sequence requires 959 $ns$ to complete the read and 1787 $ns$ to complete the write, but the corresponding 8-bit binary number only requires 28 $ns$ to complete the read and 54 $ns$ to complete the write.

\textbf{Solution:} 
We adopt a strategy where SN and UN are computed and output in segments, which are then written into the nanowires in parallel. 
During this process, we observe that in the SN coding of LD-SC, the segmented structure of SN contains multiple static repeated parts and one non-repeated part. Based on this observation, we design a new hybrid coding scheme called Pseudo-Fractal Compression (PFC). This scheme combines the static repetitive part of the segment with the binary part, aligning the repetitive part and a single bit from the binary part in time (as detailed in section 3.3). 
By integrating PFC, we redesign the multiplier so that the multiplication results are generated in segments on demand. This enables the system to efficiently produce only the required segments for subsequent processing, optimizing both storage and computational resources.
Compared to the full SN, the compression efficiency of the PFC is at least twice as high, and it rises as the length of the BN increases.




\subsubsection{Misalignment ET slowing the concurrency of vectors}

\begin{figure}[!t]
    \centering
    \includegraphics[width = 0.9\linewidth]{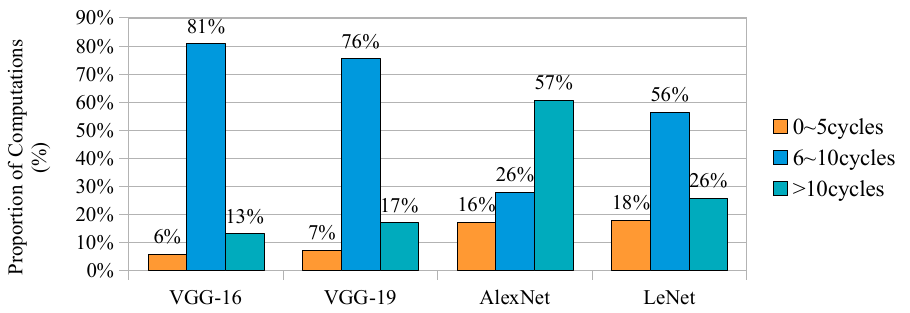}
    \caption{Proportion of computations based on fastest-to-slowest cycle wait ratio across DNN models.}
    \label{fig:wait}
\end{figure}

Since ET is widely used in LD-SC to accelerate the multiplication of individuals, then the end time of different multiplication units in vector is different.
The early completed computations need to wait for the slowest ones to finish called misalignment. 
Fig.~\ref{fig:wait} illustrates the distribution of computations across different DNN models within various waiting cycle ranges. The waiting cycles indicate the number of cycles required to synchronize the fastest and slowest computations within the same convolution operation, representing the maximum number of cycles needed for synchronization.
The horizontal axis displays different DNN models: VGG-16, VGG-19, AlexNet, and LeNet. 
The vertical axis indicates the proportion of convolution operations that require a specific number of waiting cycles to the total number of convolution operations. 
It can be observed that there is a significant waste of waiting cycles. 
Taking the VGG-16 model as an example, 80.8$\%$ of the convolution operations need to wait for 6~10 cycles, while 13.2$\%$ require more than 10 cycles, with an average waiting cycle count of $8.6$. 
The misalignment of computation delays in ``AND'' gates leads to a substantial waste of waiting cycles and a very low workload to TR.

\textbf{Solution:} To address the issue of low parallelism in hardware resources caused by synchronization, we design ``asynchronous write-in with synchronous TR'', which employs multiple stacks on nanowires where the data processed earlier is first placed into the stack. 
Once the stack is full, TRs are utilized to read out the bits as intermediate results, which are then continuously accumulated until the computation of the last dot product unit is completed, yielding the final result. 
This approach reduces the overall waiting time and achieves high parallelism in hardware resources, effectively leveraging the high throughput capability of TR.

\subsubsection{Inaccessible neighbor parts within nanowires}
As shown in Fig.~\ref{Parallel TR}, the Part.1 and Part.3 can be read in parallel by raising $\phi_0$ and $\phi_2$ to $VDD$ to turn on $MTJ_0$ and $MTJ_2$. 
Meanwhile $\phi_1$ and $\phi_3$ are set to $VSS$ (off) to prevent current flow through the other parts. 
The read current is provided from $SL$ and $BL_1$ and collected at $BLB_0$ and $BLB_2$. 
Similarly, the Part.2 and Part.4 can be read simultaneously.
However, the neighbor parts cannot be accessed simultaneously with TR because they have a shared domain, at which, the two detecting currents cannot be activated in one cell. 
The inaccessibility of neighbor parts could damage the throughput of TR by half.

\textbf{Solution:} Based on inaccessible neighbor parts with TR, we design an interleaving data placement that allows multiple vectors to time-division multiplex memory bus, keeping the memory bus consistently busy.

\subsection{Methodology}

To reduce the storage pressure caused by full stochastic sequences, PFC is proposed, which uses a hybrid coding to generate SN segments on demand.  However, in traditional architectures, the computed result sequences are forced to wait due to the synchronous design, limiting vectorization and parallelism. To address this, multiple stacks are implemented in RTM to merge result segments from different lanes using an "asynchronous write-in with synchronous TR" approach. This allows result sequences to be processed on a first-come, first-served basis, maximizing the high throughput of RTM and hardware parallelism. 
Finally, based on the characteristic that neighbor parts cannot
be accessed at the same time, we design an interleaving data placement specifically for RTM, fully utilizing the memory bus.

\section{ARCHITECTURE}
\subsection{Overview of Proposed Architecture}
\begin{figure*}[!t]
\centering
\includegraphics[width=0.9\linewidth]{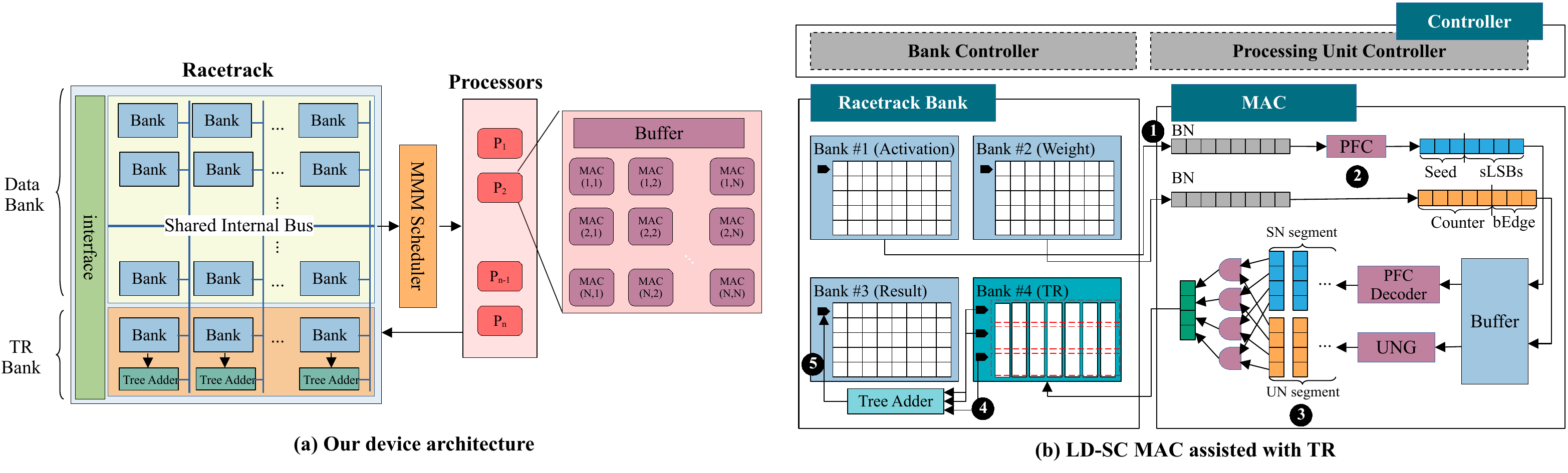}
\caption{LD-SC MAC assisted with TR Architecture. (a) shows the overview of our architecture. Different from the traditional architecture, the racetrack memory banks are divided into storage banks and TR banks, and MACs are redesigned. 
(b) shows the detailed design of our LD-SC MAC assisted with TR. The process includes \ding{182} fetch BNs; \ding{183} segment generation: encode SN to PFC and generate SN segments and UN segments on demand; \ding{184} stream computing: perform AND between SN segments and UN segments and write the result to TR banks; \ding{185} TR: execute parallel TRs on nanowires and accumulate TR results; and \ding{186} write back results.}
\label{fig:Comprehension}
\end{figure*}

Fig.~\ref{fig:Comprehension} (a) shows the internal architecture of our design.
To serve regular memory accesses, we adopt the conventional RM architecture, where banks are the top-level component and are connected by a shared internal bus to support the necessary inter-bank data transfers.
The difference is that some banks are divided into the TR unit and combined with the adders in RM to perform valid-bits collection to calculate the vector results. 
Controlled by the Matrix-Matrix Multiplication (MMM) scheduler, activations and weights stored in the banks are distributed to multiple processors for computation, with each processor equipped with multiple MAC units. The MAC units and TR unit work collaboratively to compute the vector results, and the final outcomes are written back to the result banks via the shared internal bus.

To calculate the result of one vector, the LD-SC MAC assisted with TR is designed as shown in Fig.~\ref{fig:Comprehension} (b), which consists of three racetrack banks to store the input and output, one racetrack bank to do TR for S2B conversion and the MAC to perform the dot product operations.



The computation is divided into five key steps:
\textbf{1) fetch:} In the first step, activation vectors and weight vectors stored in memory banks are transferred by the bank controller into buffers where they await processing by the computing units.
\textbf{2) segment generation:} The next step is generating SN (Stochastic Number) and UN (Unary Number) segments. This begins by comparing two binary sequences (BNs). The larger BN is converted to SN, and the smaller BN is converted to UN, simplifying the subsequent computation process (as detailed in section 3.3). SN and UN segments are generated using a quadruple consisting of the seed, LSBs (Least Significant Bits), counter, and bEdge. The seed and LSBs are determined by the Pseudo-Fractal Compression (PFC) scheme, while the counter and bEdge are based on the length of the segment.
\textbf{3) stream computing: }Since UN aggregates all '1' at the start of a sequence, the first generated UN segments can be directly streamed and written to the nanowires without performing the AND operation. The remaining SN segments, along with their corresponding UN segments, undergo the AND operation before being outputted to the nanowires. 
\textbf{4) TR:} Once the data is written into the nanowires, the TR process counts the number of valid-bits. TR accesses the valid-bits from each part of the nanowire and passes the binary results to the tree adder. 
\textbf{5) write back: }Once the tree adder completes its final operation, all valid bits from the dot-product calculation are collected. The result is then stored in the buffer as the final dot-product outcome and is subsequently transferred back into memory banks as a binary number.

In short, by leveraging the TR for valid-bits collection and streamlining segment generation with PFC, the architecture improves computational efficiency and reduces overhead.
The details of Pseudo-Fractal Compression (PFC) are further discussed in the next subsection. To support stream computing, we introduce the "asynchronous write-in with synchronous TR" in section 3.4. 
Moreover, to fully utilize the memory bus, an interleaving data placement strategy for TRs is implemented, as described in section 3.5.


\subsection{Pseudo-Fractal Compression}
In this subsection, PFC, which combines the static repetitive part of the segment with the binary part, is designed to generate segments of the stochastic sequence on demand.
Firstly, we analyze that SN hard coding provides a new SN compression scheme, PFC, and theoretically prove that this compression scheme is correct. 
Then, PFC is introduced in detail.
Finally, a computation unit that can directly perform fast multiplication using PFC while eliminating a large number of unnecessary gate-level operations is introduced.
\subsubsection{The repeatability of segments in SN}

\begin{figure}[!t]
\centering
\includegraphics[width=0.75\linewidth]{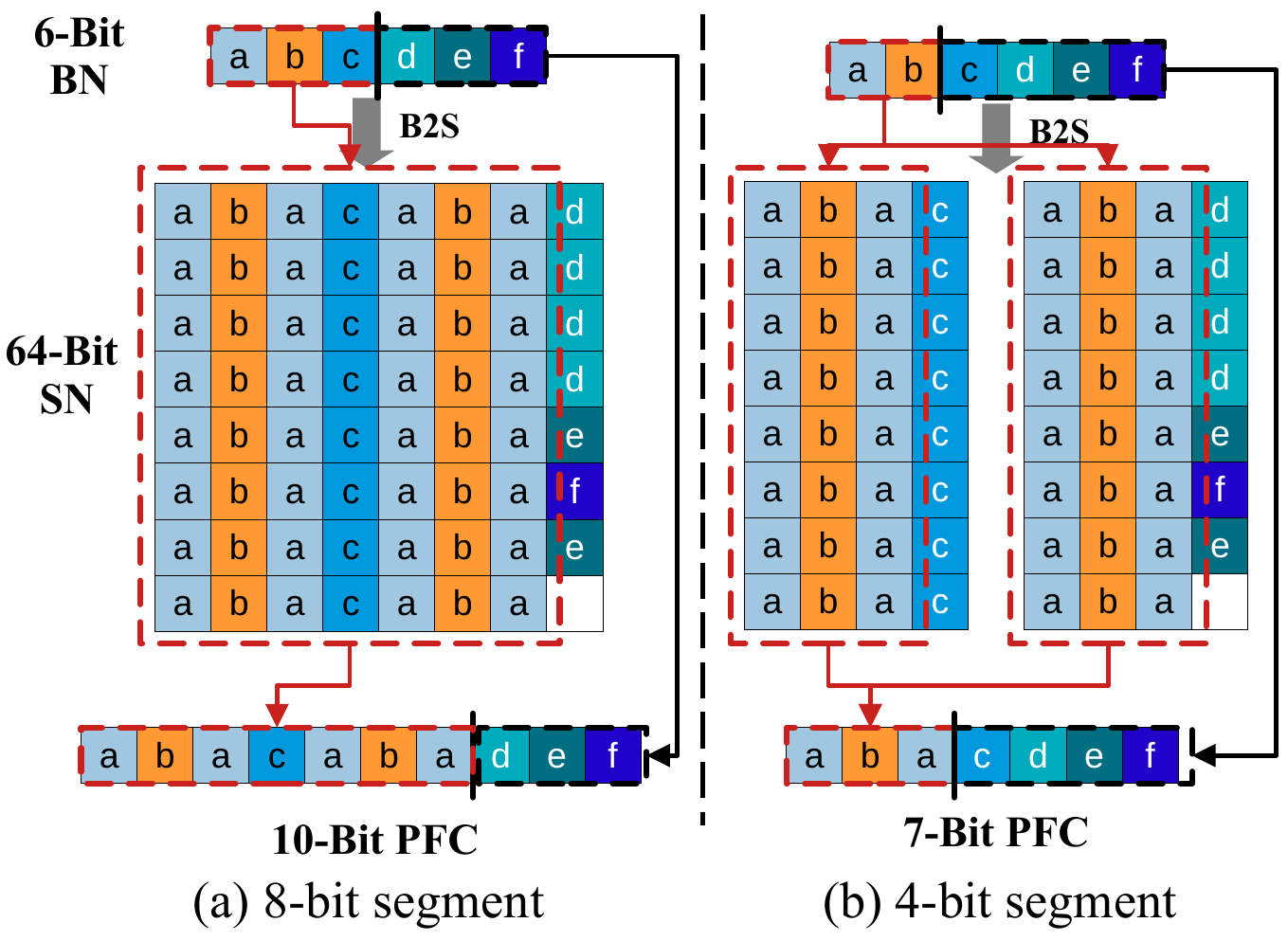}
\caption{Pseudo-Fractal compression.}
\label{SNCompression}
\end{figure}

Fig. \ref{SNCompression} shows the conversion of a 6-bit BN to a 64-bit SN.
When the 64-bit long SN is divided into 8 segments, as shown in Fig.~\ref{SNCompression}(a), each segment is 8 bits long.
It can be observed that the first 7 bits in each segment are exactly the same and the last-segment-bits (LSBs) in each segment are different.
Therefore, the generation of the SN segment can be divided into two parts, one is the static repetition part of the SN segment, and the other is the last dynamic change bit, which provides an opportunity for new segment-based generation. 
When generating SNs in segments, the first 7 bits in each segment that are all the same can be generated only once and then reused across all segments. 
The LSBs in each segment, when arranged sequentially, represent the conversion result corresponding to the final 3 bits of the 6-bit BN.
However, this method relies on the assumption that each bit in the SN can uniquely correspond to each bit in the BN; otherwise, the static repetition part between segments would not exist.

In order to prove this, we summarize the mapping equation from $n$ bits BN to $2^n$ SN in \eqref{eqn-1}. 
\begin{equation}\label{eqn-1}
S_{2^{k+1}i+2^k-1} \Longrightarrow B_k
\end{equation}
the $k$ $(k\leq n)$ means the $k$th bit in the BN and the $i$ $(i\leq 2^n/2^{k+1})$ represents the repeated times of the $k$th bit in the stochastic number and $i$ should be smaller than or equivalent to $2^n/2^{k+1}$.
For example, as shown in Fig.~\ref{SNCompression}, the fourth bit of the BN, ``d'', maps to the 7th, 23rd, 39th, and 55th positions in the SN, i.e., $k=3$. When it maps to the SN for the first time, $B_3 \Longrightarrow S_{2^{3+1} \cdot 0 + 2^3 - 1}=S_7$; when it maps to the SN for the second time, $B_3 \Longrightarrow S_{2^{3+1} \cdot 1 + 2^3 - 1}=S_{23}$; when it maps to the SN for the third time, $B_3 \Longrightarrow S_{2^{3+1} \cdot 2 + 2^3 - 1}=S_{39}$; and when it maps to the SN for the fourth time, $B_3 \Longrightarrow S_{2^{3+1} \cdot 3 + 2^3 - 1}=S_{55}$.
The current problem is that it is not clear if this formula achieves the correct conversion from the BN to the SN for all lengths of the SN and if each bit in SN can uniquely correspond to each bit in BN.
Therefore, We propose a theorem: \textbf{expression $2^{k+1}i+2^k-1$ can represent all the non-negative integers less than $2^n-1$ and uniquely correspond to each bit in BN}.

{\setlength{\parindent}{0cm}\textbf{Integrity Proof:}}

We use $I$ to represent all non-negative integers and $I$ can be divided into two parts which are even and odd, so $I$ can be represented by $2i$ and $2i+1$.
The $i$ can be still divided into two parts and then continue dividing until $i_m$.
At this point, the non-negative integer $I$ is divided into two parts, those that can be represented by the formula $2^{k+1}i+2^k-1$ and those that can be represented by $2^{m+1}{i_m}+2^{m+1}-1$.
Because the representation $2^{m+1}{i_m}+2^{m+1}-1$ can only represents the number larger than $2^{m+1}-1$, when the $n$ in \eqref{eqn-1} is less than $m$, all the numbers less than $2^{m+1}-1$ can only represent by $2^{k+1}i+2^k-1$.

{\setlength{\parindent}{0cm}\textbf{Uniqueness Demonstration:}}

So far, we have demonstrated that the \eqref{eqn-1} can be implemented in all SN\&BN conversions, but there is another question which is whether the valid-bit of SN can match a single valid bit of BN.
Assuming that a non-negative integer is $N$, and can be represented by two forms as $2^{(m+1)}i+2^m-1$ and $2^{(n+1)}i+2^n-1$, that is to say:
\begin{equation}\label{eqn-2}
2^{(m+1)}i_m+2^m-1 = 2^{(n+1)}i_n+2^n-1
\end{equation}
when $m$ is not equal to $n$, the \eqref{eqn-2} never can be equal and The same goes for $i_m$ and $i_n$.
The equation only holds if $m$ and $n$ are equal and $i_m$ and $i_n$ are equal, even if equal to 0.

In short, this subsection proves that \eqref{eqn-1} can ensure that all bits in the complete SN sequence can be mapped by the corresponding binary bit $B_k$ during the conversion process, and also proves that any bit in the SN sequence can only be uniquely mapped by its corresponding $B_k$, so our discovery of SN segment coding rules is applicable to SN sequences corresponding to any BN.
\subsubsection{Procedure of PFC}
\begin{algorithm}[b]
    \caption{PFC}
    \label{alg:AOS}
    \renewcommand{\algorithmicrequire}{\textbf{Input:}}
    \renewcommand{\algorithmicensure}{\textbf{Output:}}
    
    \begin{algorithmic}[1]
        \REQUIRE BN, Segment\_Length 
        \ENSURE PFC   
        
        \STATE\( BN \): The binary number.
        \STATE\( Segment\_Length \): The length of each segment.
        \STATE\( Seed\_Length \): The length of seed.
        \STATE\( Seed \): The repetitive\_part in all SN segments.
        \STATE\( Max\_Index \): Max index of BN bits corresponding to seed.
        \STATE\( LSBs \): All BN bits corresponding to the last dynamically changing bit of each segment of SN.
        \STATE Seed\_Length = Segment\_Length - 1
        \STATE $Seed =\{S_0,S_1,...,S_{Seed\_Length - 1}\}$

        \FOR{each element in $Seed$}
            \STATE $S_j \Longrightarrow BN_k$
             \IF{$k > Max\_Index$}
                \STATE $Max\_Index = k$
        \ENDIF
        \ENDFOR

        \STATE$ LSBs = \{ BN_{Max\_Index+1}, ..., BN_{length(BN)-1} \}$

    \STATE $PFC = \{ Seed, LSBs\}$

    \STATE \textbf{Return:} $PFC$
    \end{algorithmic}
\end{algorithm}

Since each bit in BN can be uniquely and completely mapped to each bit in SN, when BN is known, the value of each bit in SN is uniquely determined and can be realized by hard connection. 
At this time, when SN is divided into multiple segments, these segments are composed of two parts, one part is a static repetition of bits, and the other part is a dynamic change of the last bit.
PFC extracts the repetitive part in SN segments and stores them only once and then stores the remaining parts in the binary form.
Firstly, the repetitive part in each segment, referred to as the seed, is determined by the length of the segment such that $length_{seed} = length_{segment} - 1$ and obtained directly from the SN through the hard connection. 
Then, the maximum index of BN bits corresponding to the seed is determined, and all bits whose index in BN is greater than the maximum index are written into LSBs.
At this time, the seed and LSBs are spliced to form PFC.
As shown in Fig. \ref{SNCompression}, the PFC combines the repetitive 7 bits in each segment and the binary 3-bit segment LSBs into a new hybrid 10-bit coding.

\begin{figure}[!t]
\centering
\includegraphics[width=0.9\linewidth]{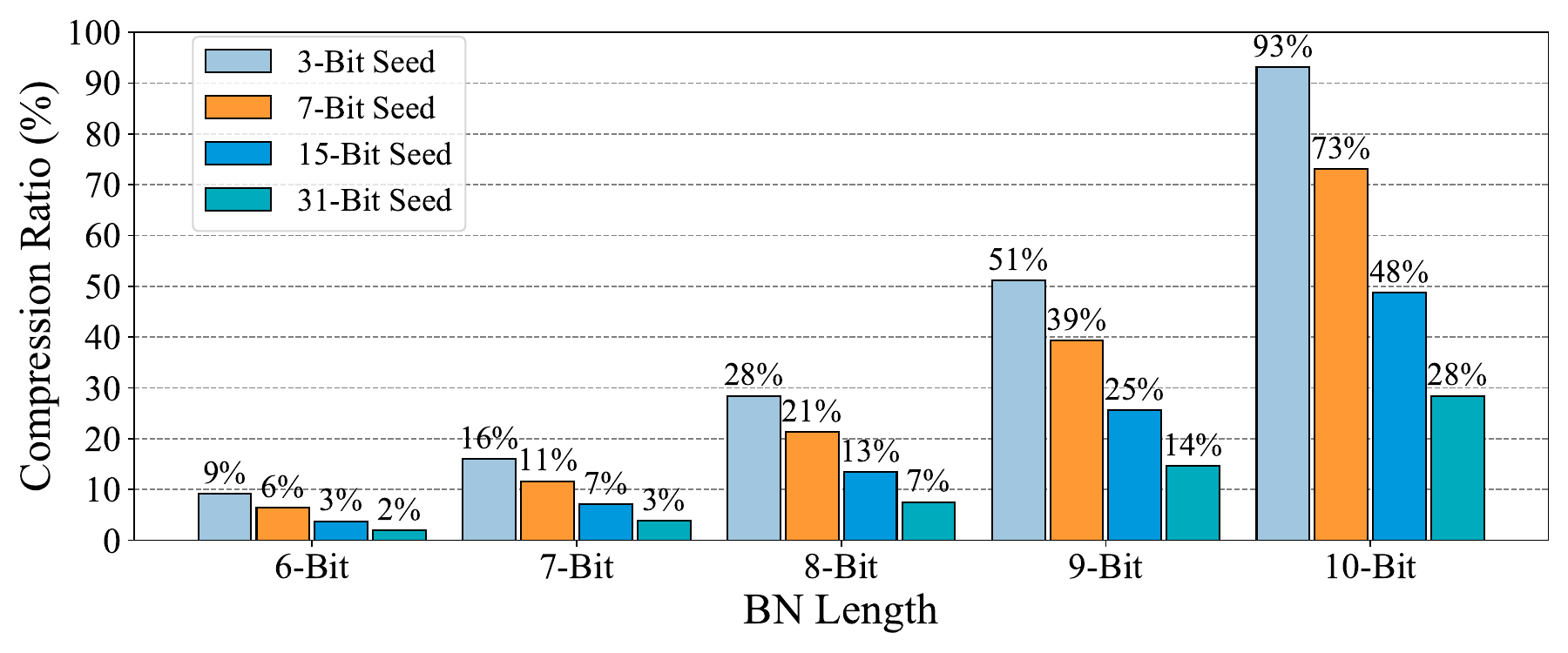}
\caption{Compression ratio of multiple seeds.}
\label{CompressionRatio}
\end{figure}

The seed length can be adjusted shown in Fig.~\ref{SNCompression}(b), when the segments are 4 bits long means that the seed is 3 bits long, which comes from the first 2 bits in the 6-bit BN.
At this time the LSBs become 4 bits, therefore, the PFC coding is 7 bits long including a 3-bit seed and 4-bit LSBs and if the seed is shorter, which means the segment is shorter, the compression efficiency is higher.
Fig.~\ref{CompressionRatio} shows the compression efficiency of PFC with multiple seeds, compared to the full SN, the compression efficiency of the PFC is at least twice as high, and it rises as the length of the BN increases.
\subsubsection{PFC-based Multiplication}
\begin{figure}[!t]
\centering
\includegraphics[width=0.5\linewidth]{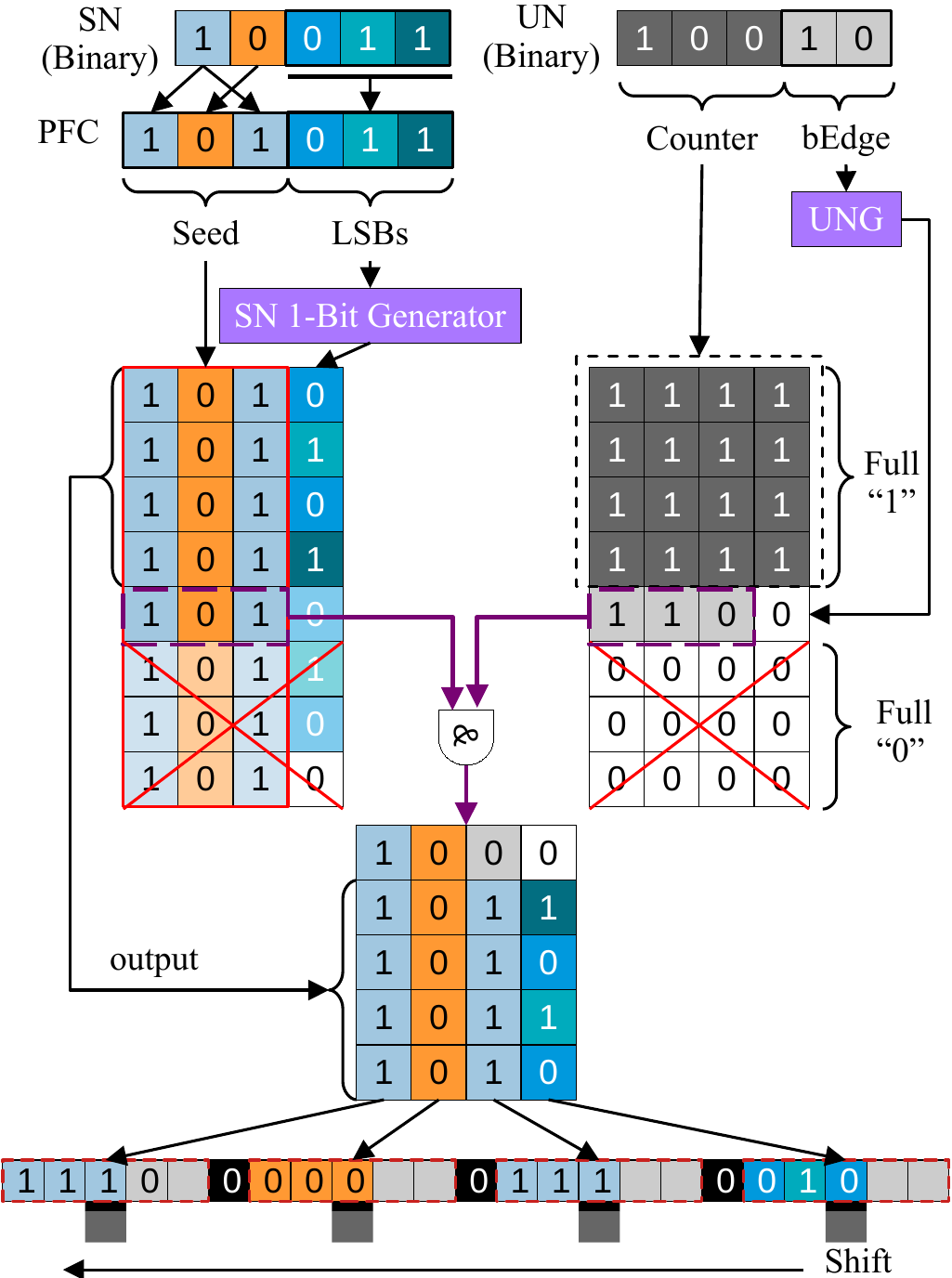}
\caption{PFC SN multiplication.}
\label{fig:5biteg}
\end{figure}

But these SNs compressed by PFC need to ensure that they can be used directly in LD-SC, otherwise re-expanding them into complete SNs will still increase the computing pressure.
We design PFC SN Multiplication (PFC-SNM) and use two 5-bit operands to analyze the process of LD-SC shown in Fig.~\ref{fig:5biteg}.
In Fig.~\ref{fig:5biteg}, the SN and UN are partitioned into 8 segments of 4 bits and it can be clearly seen that the UN has three different kinds of segments, which are full '1' segment, mixed segment, and full '0' segment.
In LD-SC, SN segments AND with the full '1' segments output SN segments themselves, and AND with the full '0' segments output full '0' segments.
Only the mixed segment needs to be sent into the AND-gate logic unit with its corresponding SN segment.

In this case, traditional LD-SC can be simplified by removing the AND operation of full '1' segments and full '0' segments.
The new LD-SC assisted with TR directly outputs SN segments into RTM with the corresponding full '1' segments in the UN and we name this computing process as output computation.
Segments of SN are generated by seed and the LSB from the SN 1-bit generator.
To support ET, we only output segments that contain '1',
so the number of generated SN segments depends on the value of the counter in the UN which controls the number of full '1' segments.
In this example, the length of each segment is 4 bits, so the value of the counter is the former 3 bits of binary UN and there are 4 full '1' segments.
We use the smaller of the two operands as UN code to reduce the number of output segments and speed up the whole calculation process.
The mixed UN segment can be converted by UNG from the binary edge (bEdge) part which includes the last bits in binary UN and then be sent into AND-gate logic units with its corresponding SN segment to do the mixed computation, which is the only AND operation in each LD-SC multiplication.
Since the last bit of a mixed segment is always '0', the UNG only needs to generate the first three bits, while the SNG only needs to generate the seed without generating the LSB to do the AND operation.
Finally, the AND result is streamed out into RTM.
If the bEdge is all '0', it means that the mixed segment is all '0' and the computing can be finished prematurely. 
Therefore, in LD-SC MAC assisted with TR, the multiplication only needs several segments of the SN and the mixed segment in the UN.
As shown in Fig.~\ref{fig:5biteg}, the streamed-out segments are stored into 4 parts on the nanowire, while the number of parts is the same as the length of the segment to ensure that in one store operation delay, a whole segment is finished storage and each part owns 5 domains between two black constant 0 domains.
After all the segments are stored in the RTM, TR operations access the parts to get the number of '1' and send these interim results into the tree adder to get the final multiplication result.

To ensure that the on-demand generation of SN does not affect the overall energy consumption and delay loss, we use a simple circuit to implement a SN 1-bit generator. 
The frequency of polling bit in LSB in the 1-bit generator of SN is actually the same as that of '1' in each bit from right to left in the accumulator. 
The accumulator accumulates '1' in each cycle, and then the frequency of '1' in the last bit from right to left is $\frac{1}{2}$, the frequency of '1' in the penultimate bit is $\frac{1}{4}$, and the frequency of the highest bit in the accumulator is the lowest. 
So in each cycle, the SN 1-bit generator outputs 1-bit LSB according to '1' of each bit in the accumulator, which is then combined with the seed to form the SN segment.
The accumulator can also be compared to the counter to determine if the output computation can be terminated.

\begin{figure}[!t]
\centering
\includegraphics[width=0.8\linewidth]{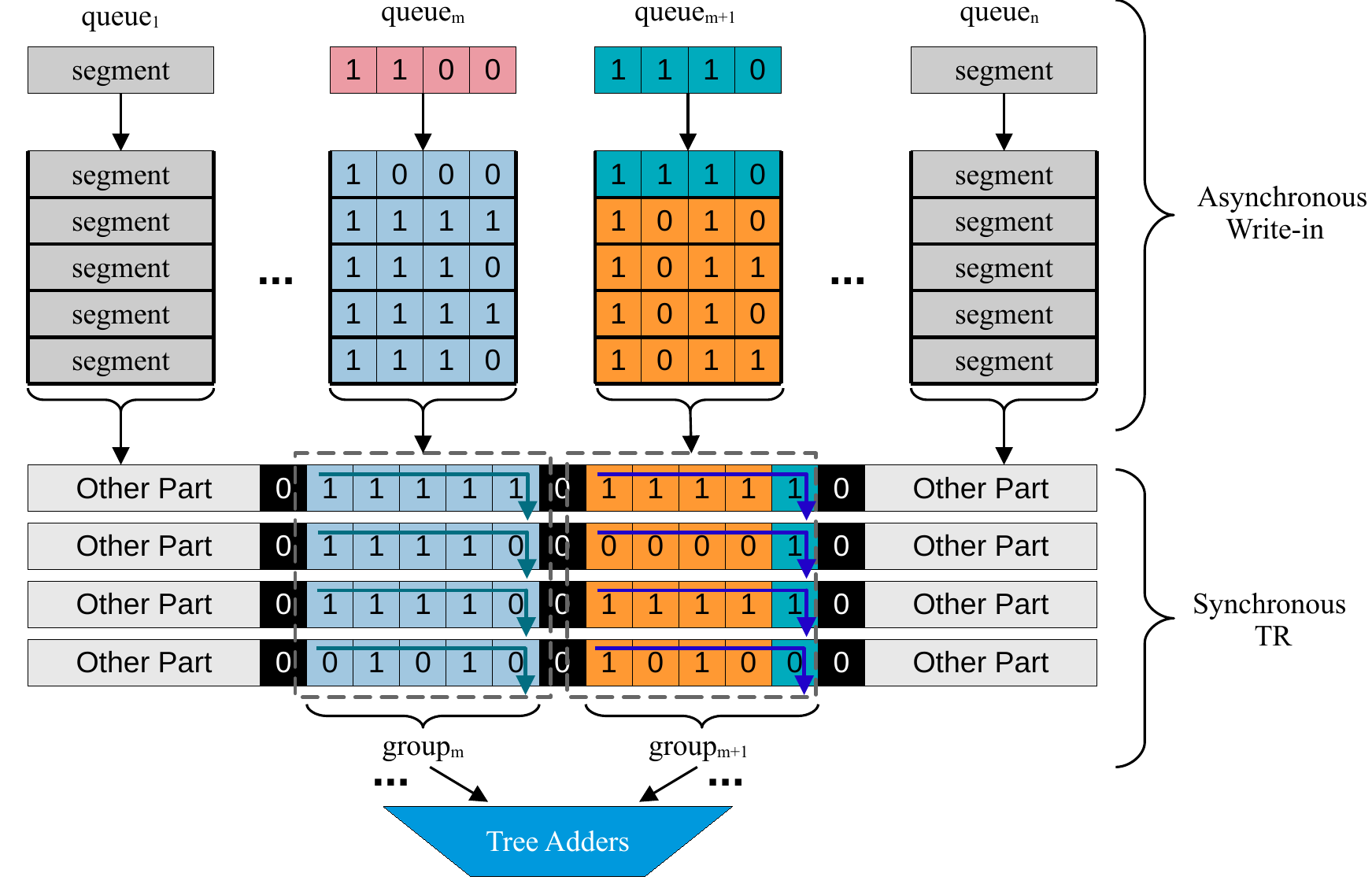}
\caption{Asynchronous Write-in with Synchronous TR. Segments are distributed across different group part queues, where the MAC computation results of the same vector share these queues. The elements of each queue are written into the corresponding parts group, and then the TR operation is executed to read out the valid bits from the nanowires.}
\label{fig:dotproduct2}
\end{figure}

\begin{figure}[!t]
\centering
\includegraphics[width=0.75\linewidth]{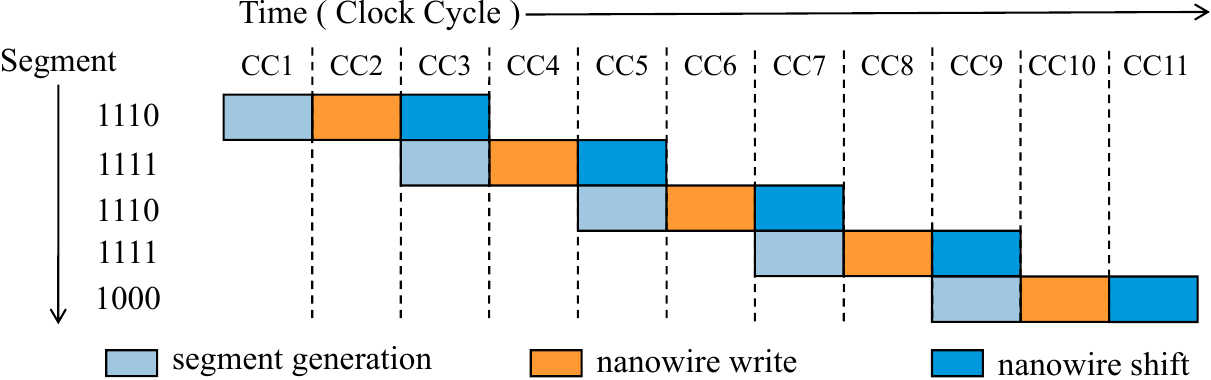}
\caption{Segments pipelined output to racetrack memory.}
\label{fig:pipeline}
\end{figure}

\subsection{Asynchronous Write-in with Synchronous TR}
The characteristic of LD-SC, as introduced in subsection 3.1, is that the effective part of the result is the number of '1', which remains the same whether the valid-bits are added after all multiplications are completed or during the calculations. 
Therefore, we can view dot product calculations not as separate multiplications, but as a process of outputting all valid-bits in the form of segments. 
Combined with the LD-SC multiplication unit based on PFC, this subsection design the ``asynchronous write-in with synchronous TR'' to optimize the multiplication operation in the same vector. 
Unlike the traditional design, which performs all multiplications before summing the results, our approach does not require waiting for all dot product units to complete. 
By utilizing multiple RTM stacks, it implements a first-come, first-served data strategy to achieve vectorized computation.



The "asynchronous write-in with synchronous TR" mechanism consists of two key components. The first component is the asynchronous write-in, which works in conjunction with the PFC-based multiplication unit. 
The parts on the nanowires are organized into multiple groups, with each group containing a number of parts corresponding to the segment length. 
This setup enables parallel segment writing into the parts within a single group. 
Each group is associated with a segment input queue, where the computed segments from different MAC outputs are distributed into available queues.
For example, as illustrated in Fig.~\ref{fig:dotproduct2}, $group\_m$ and $group\_{m+1}$ correspond to $queue\_m$ and $queue\_{m+1}$, respectively. 
In $queue\_m$, the output from one MAC unit perfectly fills the queue, while in $queue\_{m+1}$, the output from another MAC unit only partially fills the queue. 
Instead of allowing a single MAC unit to exclusively occupy $queue\_{m+1}$, segments from other MAC units continue writing into the remaining space. 
If one MAC's output exceeds the length of the queue, the excess segments can be written into the input queues of other MAC units within the same vector.
If no other MAC outputs are available to fill the incomplete $queue\_m+1$, the remaining space in the queue is padded with all-zero segments to ensure completion.

Once the parts are filled, the nanowire begins executing synchronous TR, accessing bits on the nanowires to obtain all intermediate results. These intermediate results are then sent to the tree adders for temporary storage. After all segments in a vector have been processed by TR and the binary intermediate results have been accumulated in the tree adders, the adders sum these results to produce the final dot-product output.

In our design, the entire process, which includes segment generation segment writing, and nanowire shifting, is pipelined to speed up computation until the last segment is outputted as shown in Fig.~\ref{fig:pipeline}. 
In Fig.~\ref{fig:dotproduct2}, the LD-SC MAC assisted with TR completes the dot-product in just 18 cycles, which is twice as fast as traditional LD-SC methods, due to the enhanced speed from TR operations and the asynchronous push design.

\subsection{Interleaving Placement for Full Workload on Bus}
The domain block cluster (DBC), as a storage matrix is composed of multiple (usually 8,16,32,...) nanowires, and the segments outputted from one dot-product are stored in this matrix.
First, a segment is stored in multiple parts with each bit in one part to ensure that one segment can be written to the domain in a single cycle even if it is very long. The segments of the same multiplication calculation or vector dot product calculation are clustered in the same group part as much as possible.
Now we talk about the data placement for different vectors.
In this work, it is assumed that the parallelism of the calculation unit is $2^L$, that is, the length of the segment is $2^L$, and the number of calculation units of the group vector dot product is $2m$. At this time, the number of parts calculated by the group dot product is $2m \times 2^L$.

\begin{figure}[!t]
\centering
\includegraphics[width=0.8\linewidth]{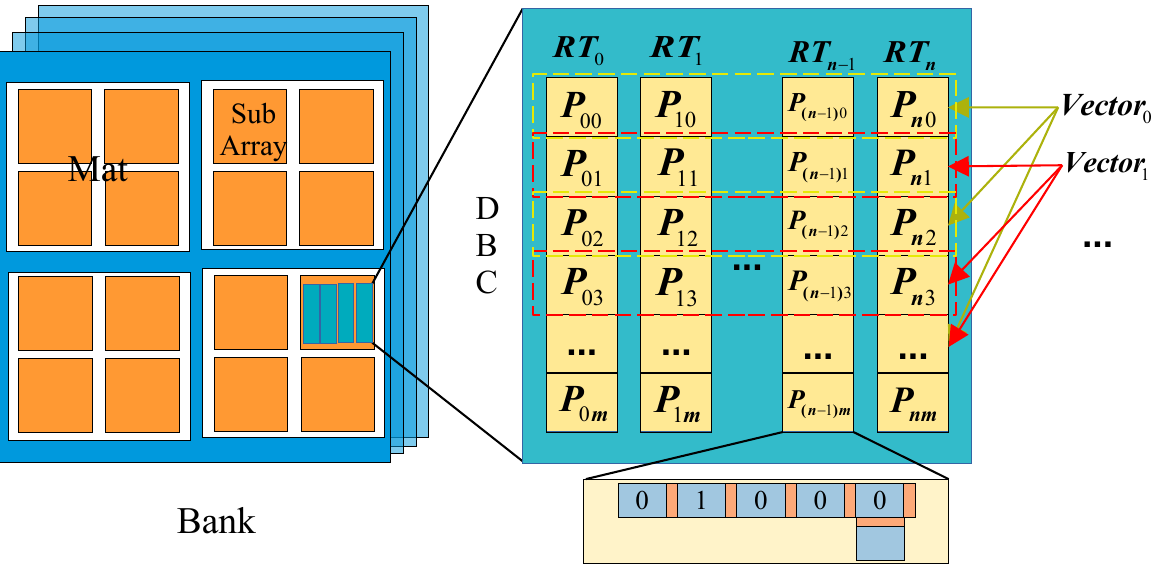}
\caption{Data placement scheme for computing. The parts of one vector are arranged at intervals.}
\label{DPSC}
\end{figure}

Because of the access pattern of TR that the neighbor parts cannot be accessed in one cycle, we design a ping-pong TR access mode shown in Fig.~\ref{DPSC}.
In this placement scheme, the parts of one vector calculation are arranged at intervals, so the TR can access all valid-bits at the same time for half of the vectors when accessing these parts, and then in the second cycle, the other vectors are read out.
At this time, the binary number of the input tree adder is $\frac{2m \times 2^L}{TRD}$ for each vector.
Therefore, in the first cycle, the valid-bit of half of the vectors are read out and sent to their corresponding tree adders to compute the final results for these vectors, which are then transmitted via the memory bus. In the next cycle, the other half of the vectors follow the same process, ensuring that the memory bus remains fully utilized.

The parts of one vector can also be continuously distributed together, which causes that TR cannot access all valid-bits at the same time for each vector and can only access half of these data.
So this placement increases the area and energy consumption of intermediate result registers because there are two TR operations to access all the parts of a vector to get the final result.
Meanwhile, the results from all the vectors can only be transmitted during half of the cycles, leading to an imbalanced load on the memory bus.
So we interleave the RTM parts among all the vectors.

\section{Reconfiguration of Parallelism and BN Length}

As Fig.~\ref{CompressionRatio} shows the seed and BN length control the compression ratio and output computation scale.
If a segment is 8 bits, the 5 bits computation example in Fig.~\ref{fig:5biteg} just needs two cycles to finish the output computation and one cycle for mixed computation, however, the nanowire needs 8 read ports to parallel access and larger tree adders to execute addition.
So, this is a situation about how to trade between the overhead of circuits and time.

\begin{table*}[h!]
\centering
\caption{Instruction set for TR-based LD-SC MAC.}
\begin{tabular}{|c|c|c|c|c|}
\hline 
    Type & IR[14:12] & Instruction & Operand & Description \\
     \hline 
     \multirow{2}{*}{Control} & 000 & TRS & - & TR-based valid-bits collection in memory starts \\
     \cline{2-5}
     & 001 & TRE & - & TR-based valid-bits collection in memory ends\\
     \hline
     \multirow{1}{*}{Compute}& 010 & TRVC & rs2, offset(rs1) & Execute TR-based valid-bits collection in memory\\
     \hline
     \multirow{2}{*}{Data Transfer} & 011 & TRW & rs2, offset(rs1) & Write the SN result to TR Bank\\
     \cline{2-5}
     & 100 & TRRW & rs2, offset(rs1) & Write the binary result calculated by TR and tree adder to a storage bank\\
     \hline
\end{tabular}
\label{table:ISA}
\end{table*}

\begin{figure}[!t]
\centering
\includegraphics[width=0.9\linewidth]{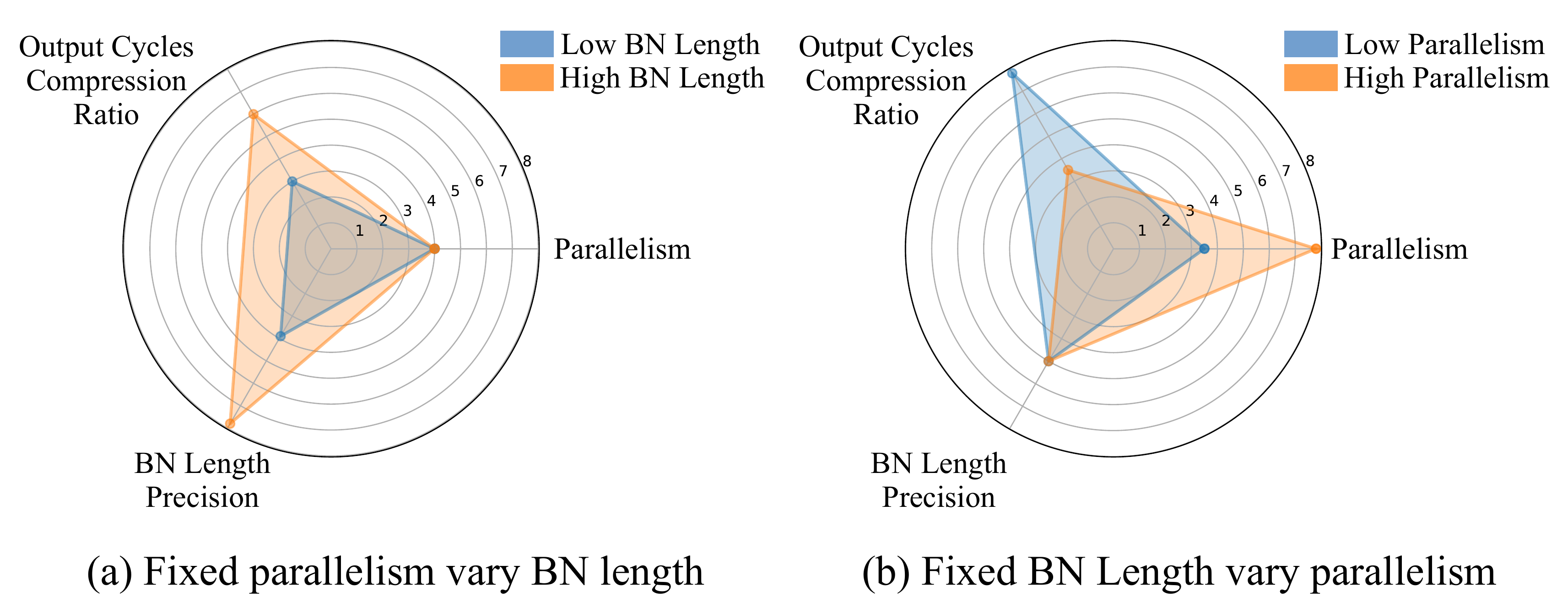}
\caption{The correlation of parallelism, output cycles, compression ratio, BN length and precision.}
\label{fig:radar}
\end{figure}

In Fig.~\ref{fig:radar}, we qualitatively analyze the relation among Parallelism (P), Output Cycles (O), Compression Ratio (R), and BN Length (Precision) (L).
Parallelism (P) represents how many bits can be parallelly outputted into RTM, that is the length of a segment; Output Cycles (O) means how many cycles are needed to finish the computation;
Compression Ratio (R) refers to the change of SN sequence length before and after coding;
BN Length (L) is the length of the BN sequence, representing the precision of the data and the longer BN Length has more bits to get higher precision of computation in decimal computation. 
As shown in Fig.~\ref{SNCompression} and Fig.~\ref{fig:5biteg}, for Output Cycles and Compression Ratio, the more cycles are needed the higher the compression ratio is, because shorter PFC SN has more LSBs and shorter seed, which will need a lot of time to generate the output computation.
When P is fixed, the longer BN means the compression ratio is higher, which needs more cycles to finish computation and more output computation overhead.
When L is fixed, the higher parallelism means the longer seed, which can accelerate output computation with less cycle output.

To vary the output computation overhead when P and L are fixed, we optimize for these two cases separately, trying to get the best configuration and the detailed experiment is shown in section 5.5.
We divide the overhead of the whole architecture into three parts: energy overhead in Computation ($E_C$), energy overhead in Racetrack ($E_R$), and energy overhead in adders ($E_A$).
$E_C$ is composed of operand converting energy, segment LSB generating energy, the energy of selectors controlling the output ways, and output energy.
$E_R$ is caused by the operations on RTM which are segments writing energy, domain walls shifting energy, and TR energy.
The tree adders and the accumulative adders' energy overhead are the source of $E_A$.
We utilize energy per operation ($E_p$) and energy-delay product ($EDP$) to evaluate the energy efficiency and speed of precision reconfigurable and parallel reconfigurable architectures, respectively.
\begin{equation}\label{eqeo}
E_p=\frac{E_C+E_R+E_A}{l_{BN}}
\end{equation}
\begin{equation}\label{eqedp}
EDP=(E_C+E_R+E_A)\times T
\end{equation}
where $l_{BN}$ is the bit length of binary numbers.

\section{Implementation}

\subsection{Programming Interface}

The data flow in the proposed LD-SC MAC architecture is different from the traditional MAC architecture. 
The traditional MAC architecture sends the calculated results into the Racetrack memory in binary form, while in our LD-SC MAC, the calculated random number sequence is directly sent to the Racetrack. 
The near-memory computing unit (TR-based valid-bits collection), which includes TR Banks and accumulation units, in the Racetrack continues to complete the subsequent calculation of the MAC and then sends the final result to the Bank, which stores the result through the bus inside the Racetrack. 
In order to ensure the correct execution of our proposed LD-SC MAC, we extend RISC-V with the Five instructions (Table \ref{table:ISA}) that provide the programming interfaces at the ISA layer.


\textbf{Control Instructions:} The newly designed control instructions, TRS and TRE, are responsible for controlling the start and end of the TR-based valid-bits collection on the memory side. 
When the processor executes the TRS instruction, it sends a command to the memory to initiate the TR operation, ensuring that the correct banks and data segments are ready for TR.
After the TR-based valid-bits collection is completed, the processor executes the TRE instruction to signal the end of the TR process, ensuring that the memory is ready for the next TR-based valid-bits collection.

\textbf{Computational Instruction:} When the TRVC instruction is executed, the stochastic sequence is first read from the memory buffer and sent to the TR Bank. After the sequence is written into the TR Bank, the TR operation is performed. Finally, the accumulated result of the TR operation is written back into the register.

\textbf{Data Transfer:} The TRW instruction writes the stochastic sequence resulting from MAC computations into the memory's buffer. This stochastic sequence is temporarily stored in the buffer, awaiting the execution of the TRVC instruction for further processing.

Also, we design a set of status registers (Table \ref{table:register}) in memory to manage banks and control the execution of the TR-based valid-bits collection. 
By setting the TRBA and numTRB, the bank in memory can be initialized as a unit that performs TR computing tasks or configured as a storage unit. 
This flexibility allows the system to flexibly allocate resources according to different workloads, improving overall computing and storage efficiency.
The distance of the TR operation is set by the TRD register, while the parallelism of the TR is determined by the PS register. 
Additionally, since the TRRW instruction requires data to be transferred through the internal bus within the memory, the sIMB register is designed to switch the state of the internal bus.
Finally, the memory system manages the available parts in the TR Bank by using a bitmap, with the BPTRP register pointing to the bitmap that tracks free parts within the TR Bank. 

\begin{table}[h]
\centering
\caption{The newly designed status registers for TR-based LD-SC MAC.}
\begin{tabular}{|c|c|}
\hline 
     Register & Description \\
     \hline 
     TRBA & \underline{TR} \underline{B}ank \underline{A}ddress \\
     \hline 
     numTRB & The \underline{num}ber of \underline{TR} \underline{B}ank \\
     \hline
     TRD & \underline{TR} \underline{D}istance \\
     \hline
     PS & The Parallel for one Segment to \\
     \hline
     sIMB & \thead{The \underline{S}tate of \underline{I}nternal \underline{M}emory \underline{B}us: 0 for interaction \\ with the CPU, 1 for internal memory operations} \\
     \hline
     BPTRP & The \underline{B}itmap \underline{P}ointer of \underline{TR} \underline{P}art\\
     \hline
\end{tabular}
\label{table:register}
\end{table}

\subsection{Matrix-matrix multiplication Scheduler}
With the development of DNNs, the scale and complexity of the models are increasing, and the computing power of a single processor has been unable to meet the performance requirements. 
In order to solve this problem, huge computing tasks should be split and assigned to multiple processors for parallel execution.

In this paper, we use SUMMA \cite{SUMMA} to schedule matrix operations among multiple processors, which gradually completes the matrix multiplication in the form of blocks through broadcast and accumulation operations.
To calculate $C = A \times B$, where matrix A is of size $m \times k$, and matrix B is of size $k \times n$, we distribute the computation across a grid of $r \times c$, Let $P_{ij}$ represent the processor in the $i-th$ row and $j-th$ column of the grid.
Each processor $P_{ij}$ is responsible for computing a submatrix of $C$.
Note that the rows of matrix $C$ are calculated from the rows of matrix $A$, and the columns of matrix $C$ are derived from the columns of matrix $B$.
Therefore, we constrain the data decomposition such that the rows of $A$ and $C$ are assigned to the same row of processors, while the columns of $B$ and $C$ are assigned to the same column of processors.
At each step, a processor row broadcasts a block of rows from $A$ to all processors in its row, while a processor column broadcasts a block of columns from $B$ to all processors in its column. Once a processor has received the corresponding row and column blocks, it performs a partial matrix multiplication to compute a portion of the submatrix of $C$.
This process is repeated for all blocks of $A$ and $B$ until the entire matrix multiplication is complete.

\section{Experiment Results}
In this section, we first compare the area overhead of generating segments between SNG and PFC.
Then we evaluate the energy consumption and latency of our proposed LD-SC MAC assisted with TR and compare these metrics with other computing units based on RTM to quantify the advantages of our proposed units.
To ensure the fairness of the comparison, we calculate the energy consumption and latency of multiplication and addition operations for computational units which are from CORUSCANT\cite{CORUSCANT}, SPIM\cite{SPI}, and DW-NN\cite{DW} in the same energy consumption and latency of read, write, shift and TR operations.
Next, we show that our model has good compatibility with ET, and can effectively use ET to achieve low energy consumption and high throughput systems. 
We compare the total energy consumption and total latency of a single classifier computation in VGG19, Alexnet, Lenet, and Resnet18 neural networks, where we are outperforming the other computational units in a single computation, and thus the total energy consumption and latency will theoretically be outperforming the other PIM architectures as well.
What's more, we evaluate the energy consumption per bit, EDP, and latency for parallelism and precision reconfiguration, which are discussed in section $4$, on VGG19.
Finally, we evaluate the accuracy of LD-SC MAC assisted with TR.
In the next subsection, we will describe the setup of the experiment.
\begin{scriptsize}
\begin{table*}[!ht]
    \centering
    \caption{Experimental setup \& RTM parameter.}
    \label{table:setup}
    \begin{tabular}{|c|c|c|c|}
        \hline
        \multicolumn{4}{|c|}{\textbf{Racetrack properties}} \\
        \hline
        Racetrack width/length/thickness & $32nm$ / $64nm$ / $2.2nm$ & Number of bits per racetrack & $256$ \\
        \hline
        Critical current density for TR & $3.5 \times 10^7 A/cm^2$ & Number of used bits per racetrack & $193$ \\
        \hline
        \multicolumn{2}{|c|}{shift / write / TR latency ($cycles$)} & \multicolumn{2}{c|}{$2$ / $2$ / $5$} \\
        \hline
        \multicolumn{2}{|c|}{shift / write / TR energy for once op ($pJ$)} & \multicolumn{2}{c|}{$0.3$ / $0.1$ / $0.175$} \\
        \hline
        \multicolumn{4}{|c|}{\textbf{Bank pool properties}} \\
        \hline
        Number of tracks per DBC (wordsize) & $32$ & Number of DBCs per bank & $256$ \\
        \hline
        Number of banks per bankpool & $2048$ & Number of bankpools per channel & $1$ \\
        \hline
        Number of channels per system & $1$ & Number of access ports per track & $33$ \\
        \hline
        \multicolumn{2}{|c|}{Addressing Scheme} & \multicolumn{2}{c|}{Channel:Bankpool:Bank:DBC:Domain} \\
        \hline
    \end{tabular}
\end{table*}
\end{scriptsize}
\subsection{Experimental Setup}
To facilitate a fair comparison, we benchmark the read, shift, and write latency and energy against those reported in DW-NN \cite{DW}. To comprehensively assess the racetrack's performance, we employ RTSim, an open-source domain wall memory simulator, which calculates all relevant latency and energy consumption using. 
All relevant RTM parameters are summarized in Table \ref{table:setup}. 
Utilizing operational data from CORUSCANT \cite{CORUSCANT}, TRD is set to 7 domains.
Specifically, each track is configured with 256 domains, out of which the TRD spans 7 domains, with 5 domains containing valid data and the remaining two domains at both ends set to a constant value of '0'. 
Given the inherent constraint that adjacent segments cannot simultaneously execute TR operations, we divide each track into 32 partitions, effectively utilizing 193 domains. Consequently, during any single TR operation, only 16 partitions within a track are accessible for data manipulation.

The segment output unit and tree adders in this work are synthesized in 45 nm technology using FreePDK45\cite{PDK} Design Compile (DC).
In our design, we use 8-bit multiplication, which corresponds to a 256-bit SN.
Due to the different segment lengths, the output logic has 6 different versions, Table \ref{table:ouput} shows the output latency of each logic version.
When the seed is composed of the first 2 bits in the SN, the output logic is the slowest version and the fastest version of the seed consisting of the first 63 bits outputs only 4 segments at worst.
In the experimental comparisons, we choose the fastest version as output logic.
In the implementation stage, considering there are both positive and negative values in neural network multiplication, the track is divided equally into two halves, storing positive and negative LD-SC segments respectively.
But in the output logic, we use unsigned computation, all the TR results of the additions in the final same-signed tree adder will have a sign bit added to distinguish between positive and negative values. These will then be fed into the last signed adder of the entire addition module.

\begin{table}[!h]
    \centering
    \caption{8-bits multiplication output logic \& tree adder power ($mW$).}
    \label{table:ouput}
    \resizebox{\columnwidth}{!}{
    \begin{tabular}{|c|c|c|c|}
        \hline
        Segment Parallelism & Seed length & Largest output times & Power \\
        \hline
        $4$ & $3$ & $64$ & $0.1249$ \\
        \hline
        $8$ & $7$ & $32$ & $0.1108$ \\
        \hline
        $16$ & $15$ & $16$ & $0.0972$ \\
        \hline
        $32$ & $31$ & $8$ & $0.0848$ \\
        \hline
        $64$ & $63$ & $4$ & $0.0702$ \\
        \hline
    \end{tabular}}
\end{table}

Cifar and ImageNet datasets for image classification tasks are used to evaluate the latency and energy performance of our proposed architecture with other PIM architectures.
To fit the scale of the datasets, Alexnet\cite{AlexNet}, VGG19\cite{VGG}, Inception-V3\cite{Inception}, Resnet-18\cite{ResNet}, Squeezenet-1.1\cite{SqueezeNet} and LeNet-5\cite{LeNet} are chosen to quantify the performance of each architecture.

\subsection{Evaluation of Latency}
In this subsection, we evaluate the latency of each kind of operation in different PIM architectures based on RTM.
We apply 5 Network models on 4 PIM architectures under the same frequency of 1000MHz to record and compare the consumed cycles shown in Table \ref{table:latency}.
In Table \ref{table:latency}, the latency changes with the scale of the network, in Lenet-5 our proposed architecture is 2.88$\times$ faster than CORUSCANT and in VGG-19 is 4.40$\times$ faster than CORUSCANT and when the scale of the network is larger, LD-SC MAC assisted with TR has higher efficiency in computation.

\begin{table}[!t]
    \centering
    \caption{Latency evaluation in different architectures and comparison of speedup.}
    \label{table:latency}
    \resizebox{\columnwidth}{!}
    {\begin{tabular}{|c|c| c| c| c|}

        \hline
        \multirow{2}*{\textbf{Networks}} & \multicolumn{4}{c|}{\textbf{Architectures (cycles)}} \\
        \cline{2-5}
        ~ & Ours & CORUSCANT & SPIM & DW-NN \\
        \hline
        LeNet-5 & $2.62E+02(\downarrow)$ & $7.54E+02$ & $3.17E+03$ & $3.42E+03$ \\
        \hline
        AlexNet & $3.17E+03$ & $1.36E+03(\downarrow)$ & $6.60E+04$ & $7.16E+04$ \\
        \hline
        SqueezeNet-1.1 & $3.48E+03(\downarrow)$ & $1.26E+04$ & $5.22E+04$ & $5.66E+04$ \\
        \hline
        ResNet-18 & $5.64E+03(\downarrow)$ & $2.22E+04$ & $1.15E+05$ & $1.24E+05$ \\
        \hline
        VGG-19 & $1.06E+06$ & $4.66E+05(\downarrow)$ & $2.27E+06$ & $2.46E+06$ \\
        \hline

    \end{tabular}}
\end{table}

One source of the acceleration is that the proposed LD-SC MAC has good compatibility with ET which removes the invalid-bit that are written to RTM.
There are a large number of data values that are very small in the classifier computation of neural networks, so when performing stochastic computation, there are many '0' in a complete stochastic sequence.
LD-SC MAC assisted with TR splits the complete sequence into several segments to generate the SN segments on demand, reducing the number of write, shift, and TR operations for invalid segments, and thus speeding up the computation.
In subsection $5.4$, we will provide a detailed description of the data distribution of a single classifier computation in networks and explain why our method has good compatibility with ET.
Meanwhile, in our LD-SC MAC, the number of valid-bit has no need to be stored into RTM waiting until all dot product units finish, but is inputted into tree adders to get the addition result of several multiplications.
However, the other three architectures finish the multiplication operation by addition, then they need to calculate the sum and carry bit by bit from LSB to MSB, this is another reason why their overall latency is higher than our proposed work.

\begin{figure*}[!t]
\centering
\includegraphics[width=0.95\linewidth]{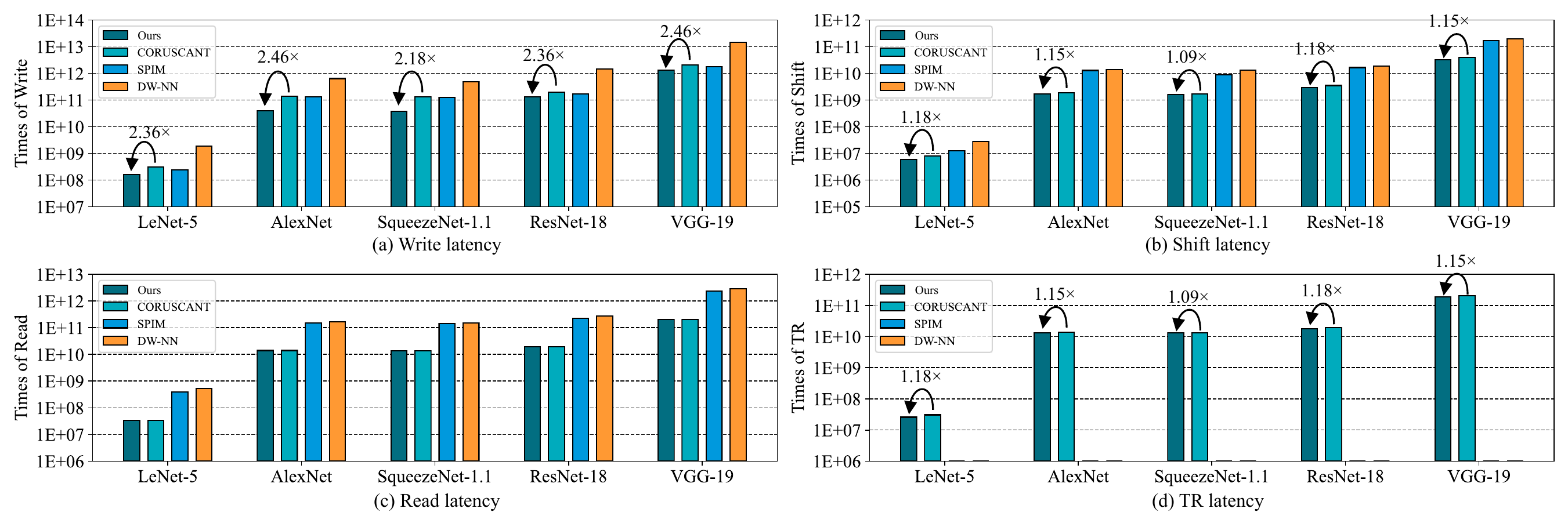}
\caption{Latency evaluation of different PIM architecture in several networks.}
\label{fig:latency}
\end{figure*}

\subsection{Evaluation of Energy}
\begin{figure*}[!t]
\centering
\includegraphics[width=0.9\linewidth, height=200 pt]{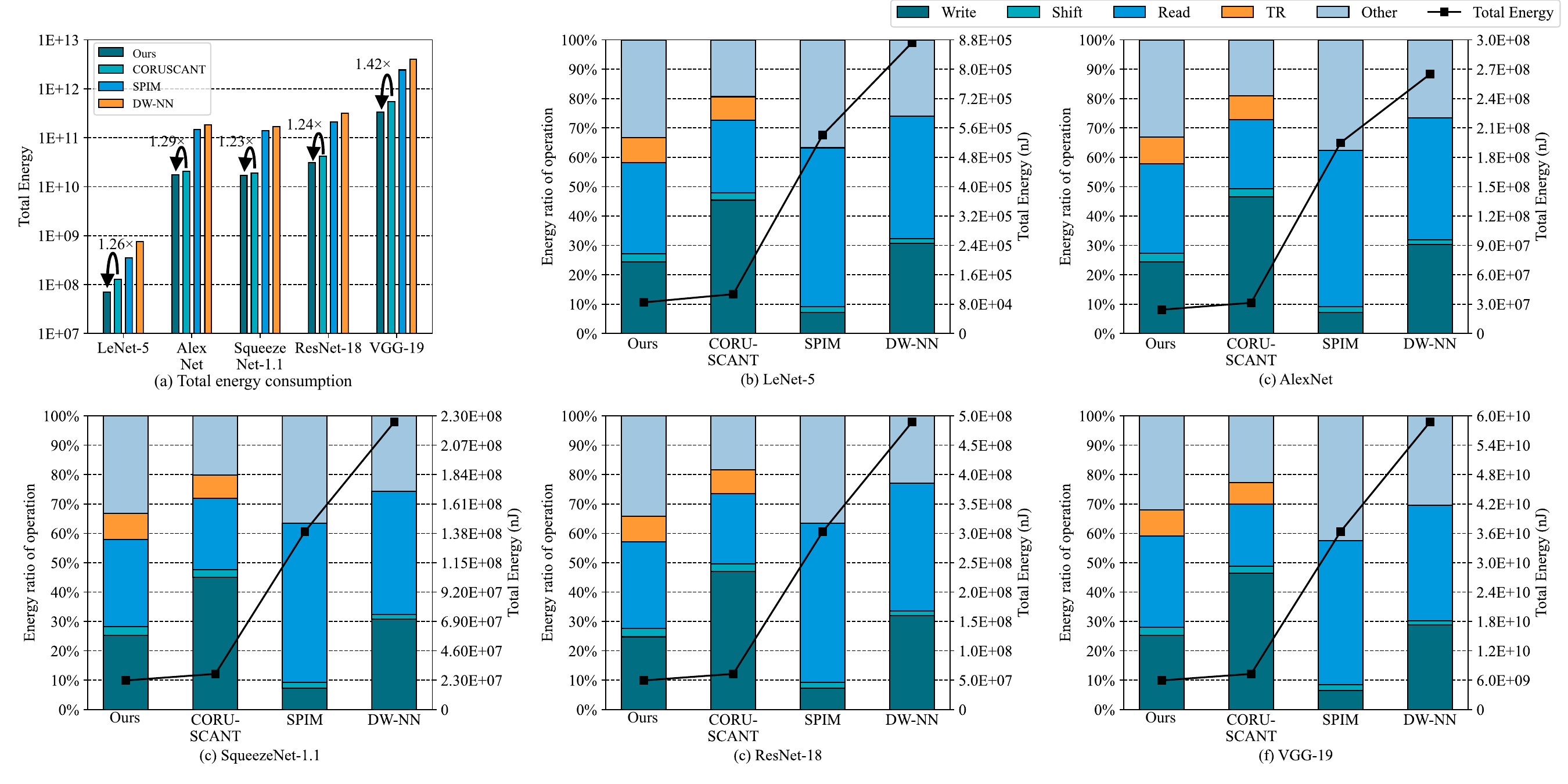}
\caption{Energy evaluation of different PIM architecture in several networks.}
\label{fig:energy}
\end{figure*}

The energy consumption in the computation of each PIM architecture comes from logic and RTM, that the logic in our proposed work consists mostly of the segments output logic and tree adders logic, and the RTM energy consumption is mainly composed of write, shift, read and TR which is the proprietary operation of LD-SC MAC assisted with TR and CORUSCANT.
The total energy consumption in the computation part is shown as Fig.~\ref{fig:energy}(a), where five different scale networks are deployed on each PIM architecture.
In small networks, e.g., our proposed work is 1.26$\times$, 6.37$\times$ and 10.3$\times$ less energy than the other three architectures, and in large networks, e.g., VGG-19, our proposed work is 1.42$\times$, 7.4$\times$ and 11.5$\times$ less energy than the other three architectures.

The low energy consumption in our proposed work is because LD-SC MAC assisted with TR has fewer write, read, and shift operations, and the main energy consumption is from the logic of outputting segments and adding TR results.
By combining multiplication and additions, which saves steps in the complete addition operations within one convolution, we ultimately reduce the total energy consumption.
The design of tree adders adding multiple multiplications results and less write operations is benefit from that the multiplications of our proposed work are more data-oriented and support the ET of sequences in the generation stage.

\subsection{Compatibility with Early-Termination}

Our work utilizes ET, which uses PFC to generate SN segments based on the size of the UN as needed. 
By combining ET with the ``asynchronous write-in with synchronous TR'', we achieve performance improvements and overcome the waste of latency for waiting. 
With ET, the UN determines the latency and energy consumption. 
In the worst-case scenario, when the UN is very large, under 64-segment-parallelism, there needs 4 cycles to generate all segments.
RTM needs 2 cycles for shifting and 2 cycles for writing, that is, in 16 cycles RTM can finish the segment writing because one part owns 5 domains, and the worst-case multiplication outputs 4 64-bit long segments which leaves the fifth domain in the 64 parts needs to be ''0'' to keep the number of '1' in this worst-case multiplication unchanged.
When the parts are filled with valid data, 64 TR operations need 5 cycles to get the results of this multiplication and send them to the tree adders waiting for 3 cycles to get the LD-SC multiplication result, which yields a total of 32 cycles and the energy consumption of one worst-case multiplication is 167.1$pJ$.
The multiplications can be parallel, the data in each part can be TR at the same time and just need another 2 cycles for the combinational logic tree adder to add two or more multiplication results together.

\begin{figure}[!t]
\centering
\includegraphics[width=0.9\linewidth]{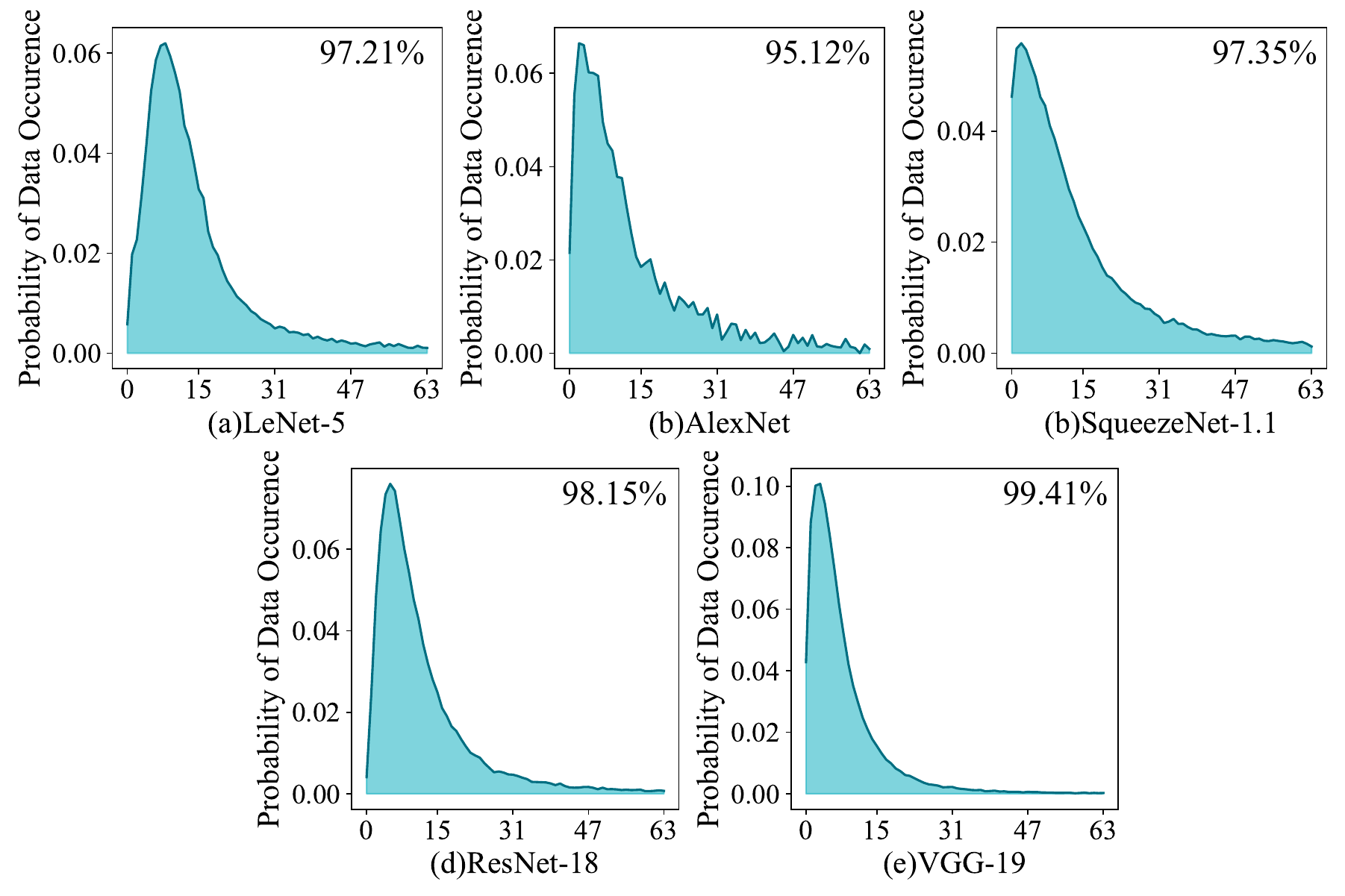}
\caption{Distribution of UN in the five DNNs. The top right corners are the percentage of the cases that only one cycle is enough for a multiplication.}
\label{fig:valuedistribution}
\end{figure}

\begin{scriptsize}
\begin{table*}[!h]
    \centering
    \caption{Operation comprasion.}
    \label{table:compare}
    \begin{tabular}{|c|c| c| c|c| c| c|c| c| c|c| c |c|}
        \hline
        \textbf{Computing Unit} & \multicolumn{3}{c|}{\textbf{Ours}} & \multicolumn{3}{c|}{\textbf{CORUSCANT}} & \multicolumn{3}{c|}{\textbf{SPIM}} & \multicolumn{3}{c |}{\textbf{DW-NN}}\\
        \hline
         & & 2 Mults & 5 Mults & & 2 Mults & 5 Mults & & 2 Mults & 5 Mults & & 2 Mults & 5 Mults\\
        \textbf{Operation} & Mult & \& & \&  & Mult & \& & \&  & Mult & \& & \&  & Mult & \& & \& \\
         & & Add & Add  & & Add & Add  & & Add & Add  & & Add & Add \\
        \hline
        \textbf{Speed (cycles)} & 32 & 32 & 34 & 64 & 90 & 90 & 149 & 198 & 328 & 163 & 217 & 357 \\
        \hline
        \textbf{Energy (pJ)} & 44.3 & 90.2 & 167.1 & 46.7 & 107.4 & 261.5 & 196 & 420 & 1101.6 & 308 & 656 & 1709.6 \\
        \hline
    \end{tabular}
\end{table*}
\end{scriptsize}

However, in network computation, in quite a few situations the worst case multiplication, that is, the multiplication operands are larger than 191 which causes the output logic to output 4 64-bit segments.
The main situation is that the output time is only once and the writing, shifting time is reduced significantly.
In order to obtain the value distribution of multiplications smaller one operand, we sampled 5 million multiplications with non-zero results in each of Alexnet, VGG19, Resnet-18, Squeezenet-1.1, and Lenet-5 and counted the number of each value appearing in multiplications.
Fig.~\ref{fig:valuedistribution} shows the distribution from ''0'' to ''255'' of multiplication smaller operands value, and all the networks distribution shows that the value of 99\% operands is between ''0'' to ''63'' which means that 64 parts with 5 domains in each part could hold the results of 5 multiplications.
The latency and energy consumption used in the worst-case multiplication now can finish 5 real multiplications.
TR in our proposed work could read 5 multiplications results using less parallelism, cycles, and energy as same as in the worst case.
In CNN computation, there are quite a few additions in average pooling layers, our proposed architecture could decrease the addition latency by parallel computers and tree adders.

Table \ref{table:compare} reports the actual speed and energy of LD-SC MAC assisted with TR, CORUSCANT, SPIM, and DW-NN for 2-operand multiplication, 2-multiplication results addition and 5-multiplication results addition in network computations.
From the data reported in Table \ref{table:compare}, our proposed LD-SC MAC assisted with TR is faster and with less energy than CORUSCANT, SPIM, and DW-NN at every aspect in actual computation because the LD-SC multiplication executes in a parallel way.
Our architecture is 2$\times$, 2.6$\times$, and 2.6$\times$ faster than CORUSCANT in 2-operand multiplication, 2-multiplication results addition and 5-multiplication results addition.
In just one multiplication, LD-SC MAC assisted with TR could put one segment on one nanowire which means that the part could fill four domains in one part like the distribution way in Fig.~\ref{fig:5biteg}.
The energy consumption is 2.4$pJ$ less than CORUSCANT and in 2-multiplication results, the addition is 17.2$pJ$ less than CORUSCANT.
In 5-multiplication results addition, 
five 64-bit long segments are written vertically into 64 nanowires and fill one part in every nanowire.
In this case, LD-SC MAC assisted with TR owns high parallelism in 5-multiplication results addition leading to the lower energy consumption which is 1.56$\times$ less than CORUSCANT and faster speed.


\subsection{Evaluation of Parallelism and Precision Reconfiguration}
In Section $4$, we have discussed the reconfiguration of LD-SC MAC assisted with TR in parallelism and precision.
The following content will give the energy consumption per bit, OPJ, and latency for two different reconfigurable cases on VGG-19.
\subsubsection{Evaluation of Parallelism Reconfiguration}
\begin{figure}[!t]
\centering
\begin{minipage}[t]{0.49\linewidth}
\centering
\includegraphics[width=\linewidth]{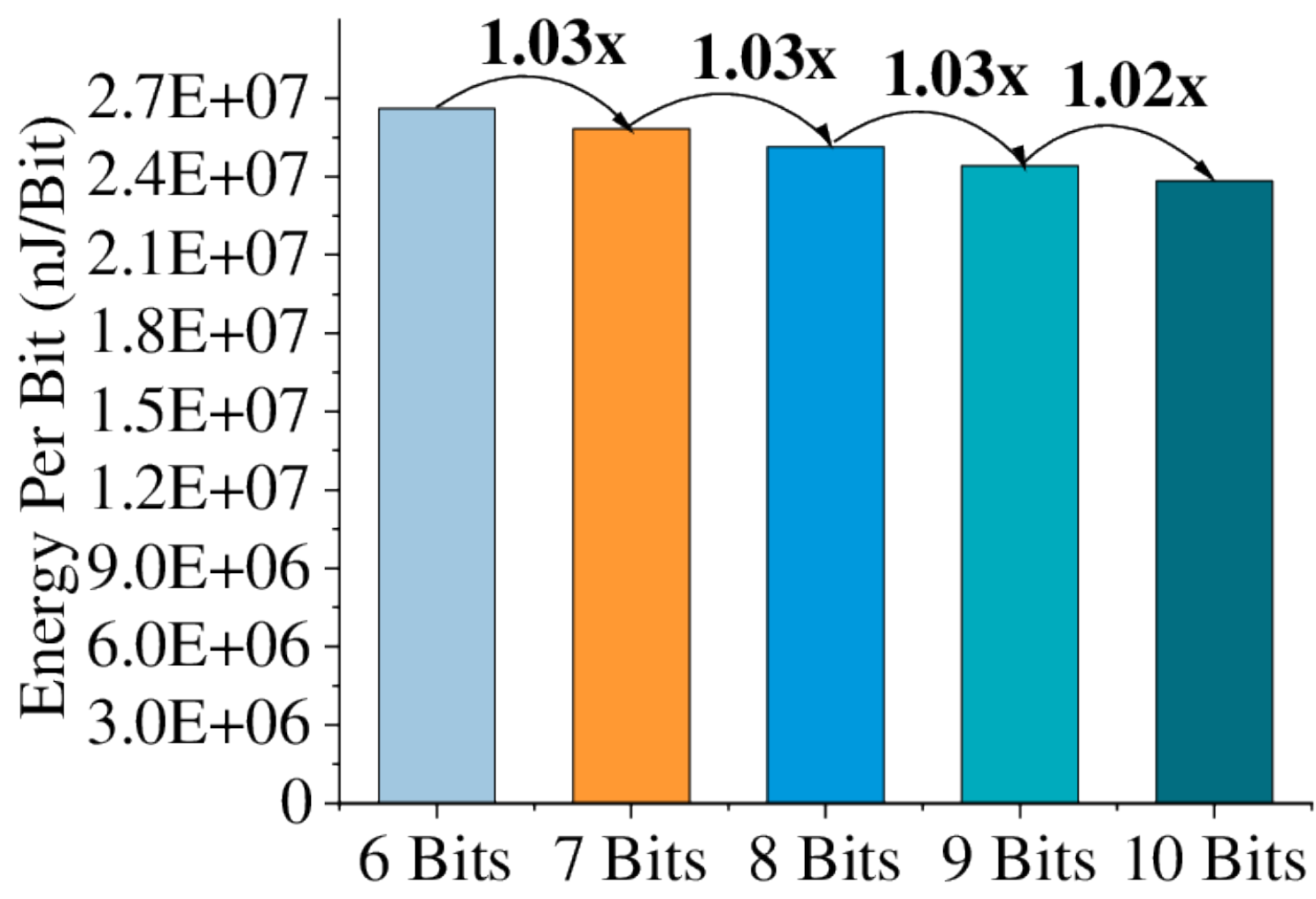}
\caption*{\fontsize{6}{5}\selectfont (a) Varying BN length on fixed parallelism}
\end{minipage}
\hfill
\begin{minipage}[t]{0.49\linewidth}
\centering
\includegraphics[width=\linewidth]{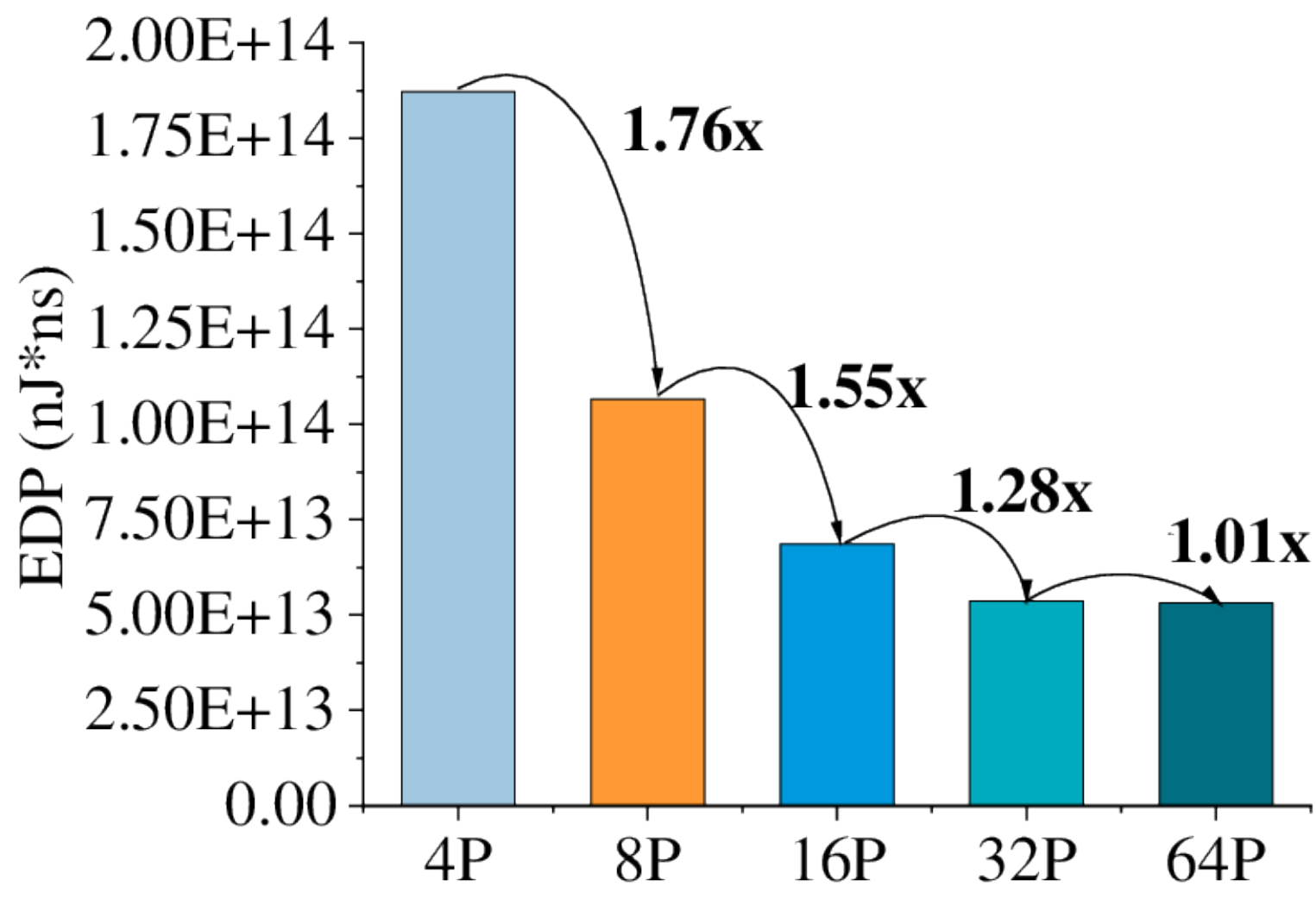}
\caption*{\fontsize{6}{5}\selectfont (b) Varying parallelism on fixed BN length}
\end{minipage}

\caption{Evaluation of reconfiguration in parallelism and BN length.}
\label{fig:reconfig}
\vspace{-0.5cm}
\end{figure}

In Fig.~\ref{fig:reconfig}(a), we take VGG-19 as an example of a setup with a parallelism of 4 at different binary data lengths.
Then the computational energy consumption per unit bit is calculated based on the \eqref{eqeo} with fixed parallelism.
Fig.~\ref{fig:reconfig}(a) shows that the energy consumption per bit decreases with the increase of BN length. This is because, in the calculation process, the main source of energy consumption is the logical output of segments and the memory writing process. 
In the case of parallelism of 4, as the binary operand increases, there will be a larger binary UN value that needs to output more segments to be written to the racetrack memory, which also increases TR operations. 
However, the increase in energy consumption of these parts is limited, and the energy consumption of a single operation is very small, so the actual increase in energy consumption is not much. 
When the length of the binary operand is 8 bits, it is only 26 $\%$ and 11$\%$ higher than the total energy consumption of 6 bits and 7 bits, but in terms of latency, the number of cycles corresponding to 6 bits, 7 bits, and 8 bits is $7.3E+03$,$2.5E+05$ and $9.3E+05$ cycles, respectively.
Therefore, in the case of fixed parallelism, the main rise is not energy consumption but latency. 
So in the reconstruction evaluation of data precise, the data precise should be selected according to the calculation, but in the parallelism selection should try to select a larger parallelism to speed up the calculation process.

\subsubsection{Evaluation of Precision Reconfiguration}
In this experiment, the data precise is fixed to 8-bit binary operands, while the parallelism is variable at five levels of 4,8,16,32 and 64. 
The experiment uses the EDP, as shown in \eqref{eqedp} to evaluate the energy efficiency and latency of the calculation. 
The latency under different parallelisms is shown in Table \ref{table:8bitlatency}. The results show that the latency increases significantly with the decrease of parallelism. 
This is because in the SN of the same length, although the lower parallelism improves the compression ratio, it also increases the number of calculations required to output segments, increasing processing time. 
On the contrary, higher parallelism can output results faster. 
In terms of energy consumption, higher parallelism will contain more '0' in the UN edge sequence, resulting in an increase in '0' in the edge sequence, thereby increasing energy consumption. 
Fig.~\ref{fig:reconfig}(b) shows the EDP under different parallelism. 
The results show that higher parallelism corresponds to lower EDP, showing better energy efficiency and latency. 
At the parallelism of 32 and 64, the EDP is close, but the parallelism of 64 is slightly better. Therefore, in the case of 8-bit binary operands, the parallelism of 64 shows the best energy efficiency and delay performance.
\begin{scriptsize}
\begin{table}[!h]
    \centering
    \caption{8-Bit Case VGG-19 Computation Latency.}
    \label{table:8bitlatency}
    \begin{tabular}{|c|c| c| c| c| c|}
        \hline
        \textbf{Parallelism} & 64-P & 32-P & 16-P & 8-P & 4-P \\
        \hline
        \textbf{Latency (Cycles)} & 105835 & 160799 & 270727  & 490583 & 930295 \\
        \hline
        \textbf{Speedup} & $\varnothing$ & 1.52$\times$ & 2.56$\times$ & 4.64$\times$ & 8.79$\times$\\
        \hline
    \end{tabular}
\end{table}
\end{scriptsize}

\subsection{Discussion of Model Accuracy}
\begin{figure}[!t]
\centering
\includegraphics[width=0.6\linewidth, height= 100 pt]{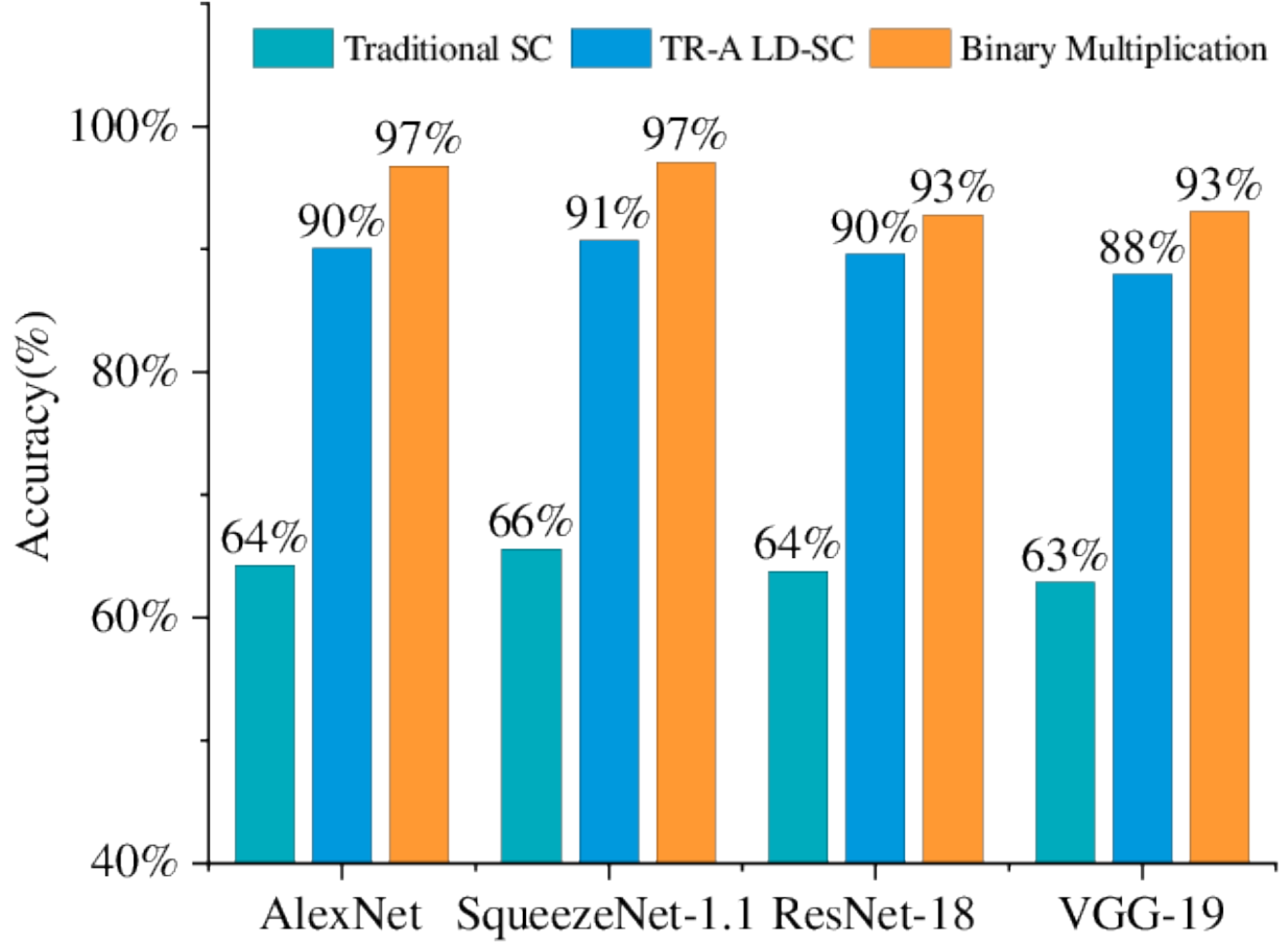}
\caption{Evaluation of accuracy between binary multiplication, traditional SC and LD-SC.}
\label{fig:precision}
\end{figure}

Currently, the APC is a significant issue affecting SC. Therefore, our research focuses primarily on addressing the APC problem, improving both parallelism and energy consumption. Compared to previous studies on LD-SC MAC accelerators, our model maintains consistent accuracy.
Therefore, we only need to compare the accuracy of different calculation methods, including traditional SC, LD-SC, and binary multiplication.
As shown in Fig.~\ref{fig:precision}, the LD-SC is slightly lower than exact multiplication but much higher than conventional SC in terms of accuracy, which means that our proposed work ensures high accuracy while significantly increasing parallelism compared to other LD-SC computations, thereby achieving low latency and low power consumption.



\section{Conclusion}
In this paper, we proposed using TR instead of APC to overcome the high energy consumption and latency issues associated with APCs.
Based on TR's access wider view of data, a low-power, low-latency neuron-architecture is designed. 
By adopting the hybrid coding PFC and ET, SN is generated by segments on demand. 
Additionally, by abandoning the traditional computing way that divides one dot-product computing into multiplication and addition steps, we designed ``asynchronous write-in with synchronous TR'' that supports computing a dot-product by executing multiplication and additions at the same time, and used a pipeline to speed up.
To solve the problem of inaccessible neighbor parts with TR, an interleaving data placement was proposed, which only requires one TR operation to access all parts of a vector to fully utilize memory bus. The evaluation shows that our design achieves lower latency and lower energy consumption compared to other PIM architectures.

\section*{Acknowledgment}
This work was supported by the National NSF (No. 62272393) of China.

\vspace{11pt}


\vfill


\begin{thebibliography}{10}
\providecommand{\url}[1]{#1}
\csname url@samestyle\endcsname
\providecommand{\newblock}{\relax}
\providecommand{\bibinfo}[2]{#2}
\providecommand{\BIBentrySTDinterwordspacing}{\spaceskip=0pt\relax}
\providecommand{\BIBentryALTinterwordstretchfactor}{4}
\providecommand{\BIBentryALTinterwordspacing}{\spaceskip=\fontdimen2\font plus
\BIBentryALTinterwordstretchfactor\fontdimen3\font minus \fontdimen4\font\relax}
\providecommand{\BIBforeignlanguage}[2]{{%
\expandafter\ifx\csname l@#1\endcsname\relax
\typeout{** WARNING: IEEEtran.bst: No hyphenation pattern has been}%
\typeout{** loaded for the language `#1'. Using the pattern for}%
\typeout{** the default language instead.}%
\else
\language=\csname l@#1\endcsname
\fi
#2}}
\providecommand{\BIBdecl}{\relax}
\BIBdecl

\bibitem{lpSC1}
M.~H. Najafi and D.~J. Lilja, ``High quality down-sampling for deterministic approaches to stochastic computing,'' \emph{IEEE Transactions on Emerging Topics in Computing}, vol.~9, no.~1, pp. 7--14, 2018.

\bibitem{lpSC2}
D.~Wang, Z.~Wang, L.~Yu, Y.~Wu, J.~Yang, K.~Mei, and J.~Wang, ``A survey of stochastic computing in energy-efficient dnns on-edge,'' in \emph{2021 IEEE Intl Conf on Parallel \& Distributed Processing with Applications, Big Data \& Cloud Computing, Sustainable Computing \& Communications, Social Computing \& Networking (ISPA/BDCloud/SocialCom/SustainCom)}.\hskip 1em plus 0.5em minus 0.4em\relax IEEE, 2021, pp. 1554--1561.

\bibitem{lpSC3}
Y.~Zhang, R.~Wang, Y.~Hu, W.~Qian, Y.~Wang, Y.~Wang, and R.~Huang, ``Accurate and energy-efficient implementation of non-linear adder in parallel stochastic computing using sorting network,'' in \emph{2020 IEEE International Symposium on Circuits and Systems (ISCAS)}.\hskip 1em plus 0.5em minus 0.4em\relax IEEE, 2020, pp. 1--5.

\bibitem{s2bsolu1}
Y.~Zhang, S.~Lin, R.~Wang, Y.~Wang, Y.~Wang, W.~Qian, and R.~Huang, ``When sorting network meets parallel bitstreams: A fault-tolerant parallel ternary neural network accelerator based on stochastic computing,'' in \emph{2020 Design, Automation \& Test in Europe Conference \& Exhibition (DATE)}.\hskip 1em plus 0.5em minus 0.4em\relax IEEE, 2020, pp. 1287--1290.

\bibitem{s2bsolu2}
W.~Li, A.~Hu, G.~Wang, N.~Xu, and G.~He, ``Low-complexity precision-scalable multiply-accumulate unit architectures for deep neural network accelerators,'' \emph{IEEE Transactions on Circuits and Systems II: Express Briefs}, vol.~70, no.~4, pp. 1610--1614, 2022.

\bibitem{s2bsolu3}
Y.~Hu, T.~Zhang, R.~Wei, M.~Li, R.~Wang, Y.~Wang, and R.~Huang, ``Accurate yet efficient stochastic computing neural acceleration with high precision residual fusion,'' in \emph{2023 Design, Automation \& Test in Europe Conference \& Exhibition (DATE)}.\hskip 1em plus 0.5em minus 0.4em\relax IEEE, 2023, pp. 1--6.

\bibitem{RTM1}
S.~Parkin, ``Racetrack memory: a high capacity, high performance, non-volatile spintronic memory,'' in \emph{2022 IEEE international memory workshop (IMW)}.\hskip 1em plus 0.5em minus 0.4em\relax IEEE, 2022, pp. 1--4.

\bibitem{RTM2}
T.-Y. Yang, X.~Peng, W.~Kang, and M.-C. Yang, ``Towards write optimization for skyrmion racetrack memory by skyrmion re-permutation,'' \emph{IEEE Transactions on Computer-Aided Design of Integrated Circuits and Systems}, 2024.

\bibitem{DNN1}
J.~Wang, Z.~Wang, and D.~Wang, ``Pseudosc: A binary approximation to stochastic computing within latent operation-space for ultra-lightweight on-edge dnns,'' in \emph{2023 60th ACM/IEEE Design Automation Conference (DAC)}.\hskip 1em plus 0.5em minus 0.4em\relax IEEE, 2023, pp. 1--6.

\bibitem{DNN2}
A.~Ren, Z.~Li, C.~Ding, Q.~Qiu, Y.~Wang, J.~Li, X.~Qian, and B.~Yuan, ``Sc-dcnn: Highly-scalable deep convolutional neural network using stochastic computing,'' \emph{ACM Sigplan Notices}, vol.~52, no.~4, pp. 405--418, 2017.

\bibitem{TRop}
K.~Roxy, S.~Ollivier, A.~Hoque, S.~Longofono, A.~K. Jones, and S.~Bhanja, ``A novel transverse read technique for domain-wall ``racetrack'' memories,'' \emph{IEEE Transactions on Nanotechnology}, vol.~19, pp. 648--652, 2020.

\bibitem{TRop2}
S.~Ollivier, D.~Kline, R.~Kawsher, R.~Melhem, S.~Banja, and A.~K. Jones, ``Leveraging transverse reads to correct alignment faults in domain wall memories,'' in \emph{2019 49th Annual IEEE/IFIP International Conference on Dependable Systems and Networks (DSN)}.\hskip 1em plus 0.5em minus 0.4em\relax IEEE, 2019, pp. 375--387.

\bibitem{SC+DNN1}
T.~Nishimura, Y.~Ichikawa, A.~Goda, N.~Misawa, C.~Matsui, and K.~Takeuchi, ``Stochastic computing-based computation-in-memory (sc cim) architecture for dnns and hierarchical evaluations of non-volatile memory error and defect tolerance,'' in \emph{2023 IEEE International Memory Workshop (IMW)}.\hskip 1em plus 0.5em minus 0.4em\relax IEEE, 2023, pp. 1--4.

\bibitem{SC+DNN2}
J.~Wang, H.~Chen, D.~Wang, K.~Mei, S.~Zhang, and X.~Fan, ``A noise-driven heterogeneous stochastic computing multiplier for heuristic precision improvement in energy-efficient dnns,'' \emph{IEEE Transactions on Computer-Aided Design of Integrated Circuits and Systems}, vol.~42, no.~2, pp. 630--643, 2022.

\bibitem{SC+RTM}
Y.~Liu, L.~Liu, F.~Lombardi, and J.~Han, ``An energy-efficient and noise-tolerant recurrent neural network using stochastic computing,'' \emph{IEEE Transactions on Very Large Scale Integration (VLSI) Systems}, vol.~27, no.~9, pp. 2213--2221, 2019.

\bibitem{DRAM-FRESH}
J.~Liu, B.~Jaiyen, R.~Veras, and O.~Mutlu, ``Raidr: Retention-aware intelligent dram refresh,'' \emph{ACM SIGARCH Computer Architecture News}, vol.~40, no.~3, pp. 1--12, 2012.

\bibitem{MAC-HIGHEN1}
H.~Zhang, D.~Chen, and S.-B. Ko, ``New flexible multiple-precision multiply-accumulate unit for deep neural network training and inference,'' \emph{IEEE Transactions on Computers}, vol.~69, no.~1, pp. 26--38, 2019.

\bibitem{MAC-HIGHEN2}
H.~Zhang, J.~He, and S.-B. Ko, ``Efficient posit multiply-accumulate unit generator for deep learning applications,'' in \emph{2019 IEEE international symposium on circuits and systems (ISCAS)}.\hskip 1em plus 0.5em minus 0.4em\relax IEEE, 2019, pp. 1--5.

\bibitem{APC-HIGH}
S.~Hu, K.~Han, and J.~Hu, ``High performance and hardware efficient stochastic computing elements for deep neural network,'' in \emph{2023 6th World Conference on Computing and Communication Technologies (WCCCT)}.\hskip 1em plus 0.5em minus 0.4em\relax IEEE, 2023, pp. 181--186.

\bibitem{ET1}
T.-H. Chen, P.~Ting, and J.~P. Hayes, ``Achieving progressive precision in stochastic computing,'' in \emph{2017 IEEE global conference on signal and information processing (GlobalSIP)}.\hskip 1em plus 0.5em minus 0.4em\relax IEEE, 2017, pp. 1320--1324.

\bibitem{ET2}
D.~Wu, R.~Yin, and J.~S. Miguel, ``Normalized stability: A cross-level design metric for early termination in stochastic computing,'' in \emph{Proceedings of the 26th Asia and South Pacific Design Automation Conference}, 2021, pp. 254--259.

\bibitem{ET3}
H.~Hsiao, J.~San~Miguel, and J.~Anderson, ``Streaming accuracy: Characterizing early termination in stochastic computing,'' in \emph{2022 27th Asia and South Pacific Design Automation Conference (ASP-DAC)}.\hskip 1em plus 0.5em minus 0.4em\relax IEEE, 2022, pp. 320--325.

\bibitem{SNG}
S.~Asadi and M.~H. Najafi, ``Late breaking results: Ldfsm: A low-cost bit-stream generator for low-discrepancy stochastic computing,'' in \emph{2020 57th ACM/IEEE Design Automation Conference (DAC)}.\hskip 1em plus 0.5em minus 0.4em\relax IEEE, 2020, pp. 1--2.

\bibitem{ARCH}
S.~Hu, K.~Han, and J.~Hu, ``High performance and hardware efficient stochastic computing elements for deep neural network,'' in \emph{2023 6th World Conference on Computing and Communication Technologies (WCCCT)}.\hskip 1em plus 0.5em minus 0.4em\relax IEEE, 2023, pp. 181--186.

\bibitem{CORUSCANT}
S.~Ollivier, S.~Longofono, P.~Dutta, J.~Hu, S.~Bhanja, and A.~K. Jones, ``Coruscant: Fast efficient processing-in-racetrack memories,'' in \emph{2022 55th IEEE/ACM International Symposium on Microarchitecture (MICRO)}.\hskip 1em plus 0.5em minus 0.4em\relax IEEE, 2022, pp. 784--798.

\bibitem{SPI}
B.~Liu, S.~Gu, M.~Chen, W.~Kang, J.~Hu, Q.~Zhuge, and E.~H.-M. Sha, ``An efficient racetrack memory-based processing-in-memory architecture for convolutional neural networks,'' in \emph{2017 IEEE International Symposium on Parallel and Distributed Processing with Applications and 2017 IEEE International Conference on Ubiquitous Computing and Communications (ISPA/IUCC)}.\hskip 1em plus 0.5em minus 0.4em\relax IEEE, 2017, pp. 383--390.

\bibitem{DW}
H.~Yu, Y.~Wang, S.~Chen, W.~Fei, C.~Weng, J.~Zhao, and Z.~Wei, ``Energy efficient in-memory machine learning for data intensive image-processing by non-volatile domain-wall memory,'' in \emph{2014 19th Asia and South Pacific Design Automation Conference (ASP-DAC)}.\hskip 1em plus 0.5em minus 0.4em\relax IEEE, 2014, pp. 191--196.

\bibitem{PDK}
C.~H. Oliveira, M.~T. Moreira, R.~A. Guazzelli, and N.~L. Calazans, ``Ascend-freepdk45: An open source standard cell library for asynchronous design,'' in \emph{2016 IEEE International Conference on Electronics, Circuits and Systems (ICECS)}.\hskip 1em plus 0.5em minus 0.4em\relax IEEE, 2016, pp. 652--655.

\bibitem{AlexNet}
A.~Krizhevsky, I.~Sutskever, and G.~E. Hinton, ``Imagenet classification with deep convolutional neural networks,'' \emph{Advances in neural information processing systems}, vol.~25, 2012.

\bibitem{VGG}
W.~Singleton and M.~El-Sharkawy, ``Effect of absolute cosine value regularization on vgg-19,'' in \emph{2021 IEEE 11th Annual Computing and Communication Workshop and Conference (CCWC)}.\hskip 1em plus 0.5em minus 0.4em\relax IEEE, 2021, pp. 0508--0516.

\bibitem{Inception}
C.~Szegedy, V.~Vanhoucke, S.~Ioffe, J.~Shlens, and Z.~Wojna, ``Rethinking the inception architecture for computer vision,'' in \emph{Proceedings of the IEEE conference on computer vision and pattern recognition}, 2016, pp. 2818--2826.

\bibitem{ResNet}
M.~D. Zeiler and R.~Fergus, ``Visualizing and understanding convolutional networks,'' in \emph{Computer Vision--ECCV 2014: 13th European Conference, Zurich, Switzerland, September 6-12, 2014, Proceedings, Part I 13}.\hskip 1em plus 0.5em minus 0.4em\relax Springer, 2014, pp. 818--833.

\bibitem{SqueezeNet}
F.~N. Iandola, ``Squeezenet: Alexnet-level accuracy with 50x fewer parameters and< 0.5 mb model size,'' \emph{arXiv preprint arXiv:1602.07360}, 2016.

\bibitem{LeNet}
Y.~LeCun, L.~Bottou, Y.~Bengio, and P.~Haffner, ``Gradient-based learning applied to document recognition,'' \emph{Proceedings of the IEEE}, vol.~86, no.~11, pp. 2278--2324, 1998.

\bibitem{SUMMA}
R.~A. Van De~Geijn and J.~Watts, ``Summa: Scalable universal matrix multiplication algorithm,'' \emph{Concurrency: Practice and Experience}, vol.~9, no.~4, pp. 255--274, 1997.

\end{thebibliography}
\end{document}